\documentclass[journal]{IEEEtran}
\IEEEoverridecommandlockouts
\usepackage[top=0.55in, bottom=0.55in, left=0.55in, right=0.55in]{geometry} 
\usepackage{graphicx} 
\usepackage{amsmath,amssymb,amsfonts,xcolor,amsthm}
\usepackage{algorithm,algorithmic}
\usepackage{stfloats}
\usepackage{colortbl}
\usepackage{cite}
\usepackage{balance}
\usepackage{epstopdf}
\usepackage{balance}
\usepackage[table]{xcolor}
\usepackage[dvipsnames]{xcolor}
\usepackage{textgreek}
\usepackage{comment}
\usepackage{graphicx}
\usepackage{caption}
\usepackage{subcaption}
\usepackage{dirtytalk}
 \usepackage[caption=false,font=footnotesize]{subfig}
\usepackage[font=footnotesize,labelfont=footnotesize]{caption}
\makeatother
\begin{document}


\title{Performance Analysis of EDMA-NOMA Enhanced V-SWAN-Assisted ISAC Systems}


\author{Arnav Hari,~\IEEEmembership{Graduate Student Member, IEEE}, Sagnik Bhattacharyya,~\IEEEmembership{Member, IEEE}, Keshav Singh,~\IEEEmembership{Senior Member, IEEE}, Meng-Lin Ku,~\IEEEmembership{Senior Member, IEEE} and Brijesh Kumbhani,~\IEEEmembership{Senior Member, IEEE}

\thanks{A. Hari is with the Institute of Communications Engineering, National Sun Yat-sen University, Kaohsiung 80424, Taiwan and also with the Department of Electrical Engineering, Indian Institute of Technology Ropar, Punjab - 140001, India (e-mail: arnav.22eez0013@iitrpr.ac.in). }

\thanks{S. Bhattacharyya and K. Singh are with the Institute of Communications Engineering, National Sun Yat-sen University, Kaohsiung 80424, Taiwan (e-mail: sagnikbhattacharyya95@gmail.com, keshav.singh@mail.nsysu.edu.tw).}

\thanks{M.-L. Ku is with the Department of Communication Engineering, National Central University, Taoyuan City 32001, Taiwan (e-mail: mlku@ce.ncu.edu.tw).}

\thanks{B. Kumbhani is with the Department of Electrical Engineering, Indian Institute of Technology Ropar, Punjab - 140001, India (e-mail: brijesh@iitrpr.ac.in).}

\vspace{-1.75em}
}
\maketitle
\begin{abstract}
This paper proposes a vertical segmented waveguide-enabled pinching antenna (V-SWAN)-assisted downlink non-orthogonal multiple access (NOMA)-integrated sensing and communication (ISAC) environmental division multiple access (EDMA) framework for simultaneously supporting multi-user communication and target sensing. In the proposed framework, multiple pinching antennas (PAs) are deployed along a segmented vertical dielectric waveguide to serve communication users (CUs), aerial users (AUs), and sensing target (ST). CUs are distributed on the ground, whereas AUs and ST are above the ground, where line-of-sight (LoS) and non-line-of-sight (NLoS) propagation conditions coexist. In particular, SWAN provides an effective platform for realizing EDMA by reducing the effective propagation distance and in-waveguide attenuation, thereby suppressing co-channel interference and enabling efficient multi-user transmission. The communication and sensing model are formulated, followed by the corresponding SINR expressions. We formulate the statistical characterizations for the combination of LoS and NLoS propagation for CUs, and LoS propagation for AUs and sensing channels for generalized multi-user, multi-PA scenarios. Based on these statistical characterizations, the segment outage and sensing outage probabilities are derived and evaluated for the generalized framework. The results show that the proposed framework outperforms the conventional H-SWAN and single vertical pinching antenna system (V-PASS) in terms of segment outage probability, sensing outage probability, and segment sum rate. Monte Carlo simulation results closely match the analytical results, validating the accuracy of the derived expressions. The results further highlight the impact of different system parameters on the performance of the proposed framework.    
\end{abstract}
\vspace{-1em}
\vspace{-2em}
\section{Introduction}
\vspace{-0.2em}
\IEEEPARstart{T}{he} growing demand for high-data-rate wireless communication systems is shifting towards millimetre-wave (mmWave) and terahertz (THz) bands for
higher bandwidth. However, these high-frequency bands also overlap with those used for radar sensing. Emerging applications such as augmented reality/virtual reality (AR/VR), Internet of Everything (IoE), and autonomous ground and aerial vehicles require both high-data-rate communication and reliable sensing capabilities. This motivates the integration of communication and sensing within the same platform. Implementing both functions as standalone systems requires redundant hardware deployment and results in high operational costs \cite{P12}. Integrated sensing and communication (ISAC) enables both functionalities by sharing the same spectrum, hardware, and signal-processing resources, thereby improving spectrum efficiency and reducing system complexity \cite{P19,A5}. Moreover, ISAC has recently attracted significant attention as a key technology for future sixth-generation (6G) networks, supporting advanced applications such as smart healthcare, environmental monitoring, and smart infrastructure \cite{ISAC1}.
\textit{Despite these advantages, ISAC operation also faces challenges due to multipath propagation, line-of-sight (LoS) blockage, clutter, and mutual interference between sensing and communication, which can significantly degrade performance even over short or moderate transmission distances \cite{ISAC2}.} 
In practical ISAC systems, multiple communication users (CUs) and sensing targets (STs) are present and need to be served simultaneously, which increases interference and makes simultaneous operation challenging. To address this problem, multiple-input multiple-output (MIMO) can be integrated with ISAC to provide diversity, multiplexing, and array gains, while also enabling spatial beamforming for multiple CUs and STs. Thus, MIMO-ISAC provides an effective approach for improving communication and sensing performance while managing interference in multi-user and multi-target scenarios \cite{ISAC3,ISAC4}. \textit{However, in conventional MIMO systems, the antennas are deployed at fixed locations. Therefore, in the presence of obstacles such as large buildings or trees, the fixed antenna positions provide limited flexibility to avoid blockage and interference, which may degrade the communication performance.}
\par To address the above issues, recent studies have explored flexible and reconfigurable antenna technologies, including movable antennas (MAs) \cite{A2}, fluid antennas (FAs) \cite{P4}, and reconfigurable intelligent surfaces (RISs) \cite{P1}. MAs mechanically adjust their antenna positions according to the users positions and varying channel conditions, while FAs employ conductive liquids in a dielectric medium to dynamically modify the antenna position and radiation characteristics. These two provide antenna position as an additional spatial degree of freedom, allowing it to be adjusted according to user requirements. Meanwhile, RISs reconfigure the wireless propagation environment by adjusting the phase shifts of the incident signals at their reflecting elements to mitigate interference and bypass obstacles. Owing to these advantages, these technologies have also found applications in ISAC systems \cite{ISAC5,ISAC6,ISAC7}. \textit{While FAs and PAs have shown performance benefits, their benefits are limited as they can only move or change their antenna positions within a few wavelengths, which has a negligible effect on the large-scale path loss. Moreover, once they are deployed in the system, they cannot be easily reconfigured.}  
\par To overcome the aforementioned  limitations, recently pinching antenna systems (PASS) have been developed which has caught the attention of researchers owing to its ability to mitigate large scale path loss effects \cite{P5}. PASS uses a dielectric waveguide as a transmission medium along which multiple PAs are deployed at different locations\cite{11612902}. The signal propagating through the waveguide is radiated into free space through these PAs, which can be placed close to the intended users to improve the propagation conditions and mitigate the impact of large-scale path loss and blockage \cite{P18}. Owing to these advantages, several recent studies have investigated the performance and design aspects of PASS \cite{P6,P7,P8,P9}. In \cite{P8}, the authors derived analytical  expressions for the outage probability and average rate for PASS while incorporating both free-space path loss and waveguide attenuation under realistic conditions. In another paper \cite{P9}, authors investigated multi-waveguide PASS for multi-user communication and proposed three transmission structures, namely waveguide multiplexing (WM), waveguide division (WD), and waveguide switching (WS).  The PA beamforming and baseband signal processing were optimized to improve the max-min fairness
(MMF) performance compared with a conventional fixed-position antenna system. In addition to this, PASS has recently attracted attention for its potential use in ISAC system \cite{ISAC8,P12,P13}. In \cite{ISAC8}  the authors derived a closed-form Cramér–Rao lower bound (CRLB) for a PASS-assisted  ISAC system and showed a significant reduction in the CRLB value compared to the conventional framework. In \cite{P13}, the author proposed a multi-waveguide PASS-assisted ISAC system, where the PA deployment and beamforming are jointly optimized to minimize the sensing CRLB while satisfying communication QoS and transmit power constraints. However, conventional horizontal PASS  is mainly designed to serve ground users and is confined to two-dimensional (2D) coverage\cite{P8,P9}, which limits its ability to support the aerial communication envisioned for sixth-generation (6G) networks \cite{P20}. In particular, unmanned aerial vehicles (UAVs) are increasingly used for applications such as urban logistics delivery, emergency services, air traffic management, and logistics support in mountainous areas. To address the aforementioned issues, vertical PASS has been introduced, where PAs are deployed along a vertical dielectric waveguide to extend the coverage from the conventional 2D plane to 3D space \cite{P22,A7}.\textit{ However, conventional PASS uses a single waveguide for all the  PAs, which can result in high in-waveguide attenuation loss and the inter-antenna radiation (IAR) effect and degraded system performance. In particular, the received  signal captured  by one  selected PA and guided via waveguide to the access point can be re-radiated by other PAs, making the received signal model more complicated and also introducing inaccuracies in the model.} 
\par
To address this issue, a segmented waveguide-enabled pinching antenna system (SWAN) is used, which divides the waveguide into multiple shorter segments having their own feed point (FP), reducing the in-waveguide distance through which the signal travels within the waveguide, as a result, the attenuation loss is also reduced \cite{P14}. In this design, the segments are isolated to prevent in-waveguide propagation across different segments, enabling the use of multiple PAs while avoiding complex electromagnetic modeling at each PA. Also, in terms of reliability, replacing  a faulty segment is more cost-effective and less disruptive than replacing the entire long waveguide. All these advantages are also applicable to ISAC system \cite{P16,P17}. In \cite{P16}, the author implemented  a hybrid segment selection and multiplexing (HSSM) protocol to provide favorable performance with lower  hardware cost. Similarly, in \cite{P17}, to balance performance and complexity, the authors proposed three transceiver protocols:\textit{segment selection}  which selects a single segment for signal transceiving, \textit{segment aggregation} which combines signals from all segments using a single RF chain, and \textit{segment multiplexing} which jointly processes the signals from all segments using individual RF chains.     
\par
On the other hand, the increasing demand for higher data rates has drawn attention to multiple access (MA) techniques such as non-orthogonal multiple access (NOMA), space division multiple access (SDMA), and orthogonal multiple access (OMA). However, these techniques are limited to conventional domains such as time, frequency, space, code, and power. Motivated by this, \cite{A4} proposed an environmental division multiple access technique (EDMA), in which key propagation characteristics, such as LoS blockage and large-scale path loss, are exploited to suppress co-channel interference and enable efficient multiple access (MA) transmission. However, traditional waveguides suffer often from  in-waveguide losses. To address this, SWAN provides an efficient platform for EDMA by reducing effective propagation distance, reducing  in-waveguide attenuation, and suppressing co-channel interference, thereby enhancing overall system performance. \textit{Although SWAN-based EDMA effectively mitigates the multi-user interference through environmental separation and waveguide segmentation, existing studies mainly focus on single-user transmission within each segment\cite{A4}. Practically, multiple users can coexist within the same segment, which makes single-user transmission inefficient in terms of spectrum utilization. So, simultaneous transmission of multi-users within each segment is essential for improving system efficiency. For this, integrating multiple access techniques like NOMA into EDMA becomes a promising solution for serving multiple users simultaneously \cite{S7}.}
\par Based on the literature survey, only the authors in \cite{P22} have recently investigated the V-PASS architecture, accounting for the combined effects of in-waveguide and free-space propagation losses. However, their study is limited to UAV communication involving a single waveguide and a single PA. Moreover, as discussed earlier, a single-waveguide architecture can suffer from considerable in-waveguide attenuation. This motivates the development of the V-SWAN architecture, where multiple waveguides and multiple PAs can reduce the effective in-waveguide propagation distance and thereby mitigate attenuation while providing greater spatial flexibility for serving distributed users and supporting sensing functionalities. To address this research gap, we propose a V-SWAN-assisted downlink NOMA-ISAC EDMA framework to simultaneously support multi-user communication and target sensing. The \textbf{main } contributions can be highlighted as follows:
\begin{itemize}
    \item \textbf{Introduction to V-SWAN Assisted downlink NOMA-ISAC EDMA Framework:} We propose a generalised V-SWAN-assisted downlink NOMA-ISAC EDMA framework, where multiple PAs are deployed in each segment to serve multiple users and ST simultaneously. The users are categorised as CUs and  aerial users (AUs). The CUs are distributed on the ground, so there are heterogeneous propagation conditions with both  LoS and non-line-of-sight (NLoS) links, whereas AUs and ST are distributed above ground, and for these, LoS conditions are considered. Based on this framework, a generalised received signal model for communication and sensing is developed, followed by the corresponding signal-to-interference-plus-noise ratio (SINR) expression.
    \item \textbf{Multi-user Multi-PA Performance Evaluation:} The proposed framework is generalised for multi-user and multi-PA scenarios. Since each segment serves multiple users, the overall segment performance is evaluated in terms of segment outage probability and segment sum rate. Sensing outage probability is evaluated for a single target. Further, these performance metrics are compared with conventional H-SWAN and V-PASS frameworks.
    \item \textbf{ Segment Outage Analysis:} Users and the target are distributed in different environments, where CUs are distributed on the ground, and AUs and ST are above the ground, their effective channel is different due to the superposition of different PA contributions with in-waveguide attenuation, phase variation, and user-dependent wireless communication.
    \begin{itemize}
        \item \textbf{ Communication  Outage:} The CU channel consists of both LoS and NLoS components, whereas the AU channel is LoS. In both cases, the effective channel is approximated by a moment-matched complex random variable with a nonzero mean, and channel power follows a non-central chi-square distribution. Based on this characterization, the user outage probability is evaluated using numerical integration, which is further used to calculate the segment outage probability.
 
        \item \textbf{Sensing Outage:} The ST channel is LoS, and the forward channel is approximated using moment matching, while the backward path is characterised by conditioning on the ST location and the selected receiving PA. Thus, the backward channel is deterministic, and its PDF and CDF are obtained accordingly. Based on this, the sensing outage is evaluated using numerical integration. 
    \end{itemize}  
     \item \textbf{Performance Comparison:} Extensive numerical results are presented to validate the analytical expression obtained from the proposed VSWAN-assisted downlink NOMA-ISAC EDMA framework for outage analysis, varying in-waveguide attenuation, different user distributions, power allocation, and sensing-related parameters, which affect the performance of communication and sensing. The obtained results outperform the existing conventional PASS framework,  and the results of the analytical expression closely match with  Monte Carlo simulation, confirming the accuracy of the proposed method.
\end{itemize}
 \begin{table}[t]
 \caption{ List of Notation and Mathematical Operations}
\label{table:1}
\centering
 {\rowcolors{3}{white!100!white!100}{white!100!white!40}
\begin{tabular}{|m{1.95cm}|p{6.45cm}|}
\hline
 \rowcolor{lightgray!80}\textbf{Symbols} & \hspace{20mm} \textbf{Description} \\
\hline
\rowcolor{Lavender!20}\multicolumn{2}{|c|}{\bf General System Parameters} \\
\hline
$X,Y,D$ & Serving area length, width and height, respectively\\
$L$ & Total number of segments\\
$A_l$,$F_l$ & Denotes $l^{th}$ segment and FP of $l^{th}$ segment \\
$A_{l,seg}$ & Segment Length of $A_l$\\
$M$ & Total number of PAs in each segment \\
$\kappa$ & In-waveguide attenuation coefficient\\
$\eta\mathrm{eff}$ & Effective refractive index\\
$c,f_c$ & Speed of light and carrier frequency, respectively \\
${\boldsymbol{\Omega}_{F_l}}$, ${\boldsymbol{\Omega}_{P_{l,m}}}$ & Coordinates of FP of $A_l$ and $m^{th}$ PA of $A_l$, respectively   \\
$\mathcal{P}_{l_{m}}$& Transmit Power of $m^{th}$ PA of $A_l$ \\
$\psi_l$ & ISAC signal of $A_l$\\
$ h_{l,m},{\textbf{h}}_{l}$ & In-waveguide channel between $F_l$ and $m^{th}$ PA and channel vector of $A_l$, respectively\\
\hline
\rowcolor{Apricot!20}\multicolumn{2}{|c|}{\bf Communication  Parameters} \\
\hline
$h^{q,m}_{l,k}$ & Channel between $m^{th}$ PA of $A_q$  and $k^{th}$ the user of $A_l$ \\ 
$ \textbf{h}^{q}_{l,k}$ & Channel vector between $M$ PAs of $q^{th}$  and $k^{th}$ user of $A_l$\\
$\theta_{q,m,l,k}$ & LoS indicator between $m^{th}$ PA of $A_q$  and $k^{th}$ the user of $A_l$ \\
$ h_{l,k}^{\mathrm{eff},q}$ & Effective channel between $F_q$ of  $A_q$ and $k^{th}$ user of $A_l$\\ 
$\nu$ & Path loss factor\\
$\phi$ & LoS blockage parameter\\
$ g_{q,m,l,k}$ & Rayleigh fading coefficient  between  $m^{th}$ PA of $A_q$ and $k^{th}$  user of $A_l$ \\
$K_{\rm R}$ & Rician factor \\
$K$ & Total No. of users in each segment \\
$\boldsymbol{\Omega}_{CU_k}$, $\boldsymbol{\Omega}_{AU_{l,k}}$ & Coordinates of $k^{th}$ CU and $k^{th}$ AU of $A_l$ respectively\\ 
$\alpha$ & Power allocation factor for communication \\
$b_{l,k} $ & NOMA power-allocation coefficient of $k^{th}$ user of $A_l$ \\
$n_{l,k}$ & Background noise at $k^{th}$ users of $A_l$\\
\hline
\rowcolor{Cyan!20}\multicolumn{2}{|c|}{\bf Sensing Parameters} \\
\hline
$1-\alpha$ & Power allocation factor for sensing\\
$\boldsymbol{\Omega}_{T_l}$ & Coordinates of ST of $A_l$\\ 
$\sigma_T$  & Radar cross section (RCS) of the ST\\
$h^{q,m}_{l,T}$ &  Forward channel between $m^{th}$ PA of $A_q$  and ST of $A_l$ \\ 
$ \textbf{h}^{q}_{l,T}$ &  Forward channel vector between $M$ PAs of $A_q$ and ST of $A_l$\\
$ \check{h}_{l,sen}$ & Effective forward channel between FP and ST of $A_l$\\ 
$\hat{P}_{l,m}$ & Selected PA for reception of reflected signal\\ 
$\widetilde{h}_{l,m}$ & Backward channel between $m^{th}$ PA and $F_l$ of $A_l$ \\
$\widetilde{h}_{T,l}$ & Backward channel between ST and selected  $\hat{P}_{l,m}$ PA of $A_l$ \\
$n_{s,l}$ & Background noise at AP for sensing\\
$\beta$ & Spectrum allocation factor\\
\hline
\rowcolor{ SeaGreen!20}\multicolumn{2}{|c|}{\bf Mathematical Operators} \\
\hline
$j, ||\cdot||, |\cdot|$ & Imaginary unit $\left(\sqrt{-1}\right)$, Euclidean Norm, Absolute value, respectively\\
$E[\cdot]$& Expectation operator\\

\hline
\end{tabular}}
\end{table}
\par
\textbf{Notations Used:} The notations and mathematical operators used in this paper are listed  in Table \ref{table:1}
\par
\textbf{Organization of the paper:} The remainder of the paper is organised as follows. Section \ref{sec:system_model} presents the system model and SINR formulations. Section \ref{sec: performance analysis} provides channel characterization and the outage analysis of the multi-user and multi-PA scenario. Section \ref{sec:result_discussion} presents the analytical and simulation results, and Section \ref{sec:conclusion} concludes the paper. 

\begin{figure}[t]
    \centering
    \includegraphics[scale=0.25]{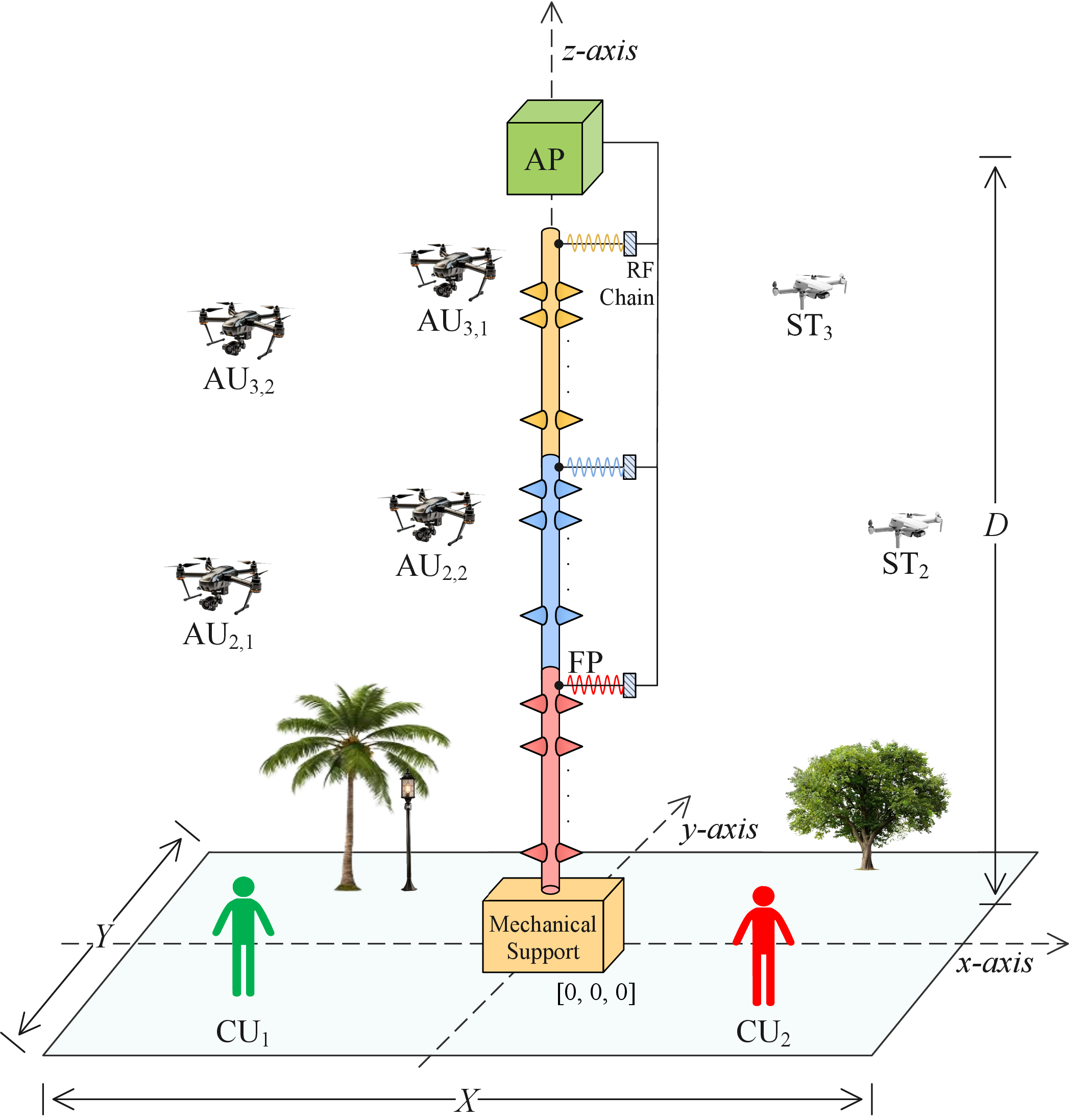}
    \caption{An illustration of the proposed V-SWAN assisted downlink NOMA-ISAC EDMA framework.}
    \label{fig:scenerio}
\end{figure}

\vspace{-1em}
\section{System Model}
\label{sec:system_model}
Let us consider a V-SWAN-assisted downlink NOMA-ISAC EDMA framework as illustrated in \figurename{~\ref{fig:scenerio}}. The complete service region of the proposed framework is given by $X\times Y\times D$, where $X$ and $Y$ denote the length and width of the considered area, respectively, while $D$ denotes the height of the waveguide. V-SWAN is mounted on a mechanical support and extends vertically along the $z$-axis. The system consists of $L$ non-overlapping segments denoted by $A_l$, where $l\in\{1,2,\ldots,L\}$ with each segment having equal length given by $A_{l,seg}=\frac{D}{L}$, equipped with a dedicated feed point (FP) and is connected to a single access point (AP) through a dedicated radio-frequency (RF) chain, through which the RF signal is injected into the waveguide. It is also equipped with $M$ PAs that radiate the guided RF signal into free space. Let $F_l$ denote the FP, while $M$ is the number of PAs randomly distributed within each segment $A_l$. Cartesian coordinates of  $F_l$ are given as ${\boldsymbol{\Omega}_{F_l}}=[0,0,Z_{F_l}]$ and, for simplicity, FP is assumed to be located at the top of each segment. PAs are represented as $ P_l=\left\{P_{l,1},P_{l,2},\ldots, P_{l,M}\right\}$ with $m\in\{1,2,\ldots,M\}$, and its  Cartesian coordinates are given by $ {\boldsymbol{\Omega}_{P_{l,m}}}= [0,0,Z_{P_{l,m}}]$, where $Z_{P_{l,m}}\in\left((l-1)A_{l,seg},lA_{l,seg}\right)$.

\par Without loss of generality, the lowest segment (l=1) only  serves the CUs denoted as $\left\{CU_{1},\ldots, CU_{K}\right\}$ and randomly distributed within the segment region. In this work, owing to mathematical tractability, we considered the case of $K=2$ in each segment. The position of the $k^{\rm th}$ CU is denoted by $\boldsymbol{\Omega}_{\mathrm{CU}_{k}}=[X_{\mathrm{CU}_{k}},\,Y_{\mathrm{CU}_{k}},\,0]$, where $X_{\mathrm{CU}_{k}}\in[\frac{-X}{2},\frac{X}{2}]$, and $Y_{\mathrm{CU}_{k}}\in[\frac{-Y}{2},\frac{Y}{2}]$ . The other upper segments ($l\neq1$ ) serve AUs denoted as $\left\{AU_{l,1},\ldots, AU_{l,K}\right\}$ which are  randomly distributed within the corresponding  segment region. The position of the $k^{\rm th}$ AU in the $A_l$ segment is denoted by $\boldsymbol{\Omega}_{\mathrm{AU}_{l,k}}=[X_{\mathrm{AU}_{l,k}},\,Y_{\mathrm{AU}_{l,k}},\,Z_{\mathrm{AU}_{l,k}}]$, where $X_{\mathrm{AU}_{l,k}}\in[\frac{-X}{2},\frac{X}{2}]$, $Y_{\mathrm{AU}_{l,k}}\in[\frac{-Y}{2},\frac{Y}{2}]$, and $Z_{\mathrm{AU}_{l,k}}\in\left((l-1)A_{l,seg},lA_{l,seg}\right)$,   and the single ST is denoted by ${T_l}$ and position is denoted by  $\boldsymbol{\Omega}_{T_l}=[X_{T_l},\,Y_{T_l},\,Z_{T_l}]$, where $X_{\mathrm{T}_l}\in\left[\frac{-X}{2},\frac{X}{2}\right]$, $Y_{\mathrm{T}_l}\in[\frac{-Y}{2},\frac{Y}{2}]$, and $Z_{\mathrm{T}_l}\in\left((l-1)A_{l,seg},lA_{l,seg}\right)$ as shown in \figurename{~\ref{fig:scenerio}}.
\par To capture diverse propagation environments, different channel conditions are considered for the CUs, AUs, and STs. Based on the above system model, the corresponding channel model is presented in the following sub-section.
\vspace{-0.75em}
\subsection{ Channel Model}
The system architecture described in the previous section enables signal transmission from the AP to either a CU, AU or an ST through a combination of waveguide and wireless propagation. Accordingly, the overall propagation channel is categorized as follows:

\subsubsection{\textbf{In-Waveguide Channel}} 
The in-waveguide channel characterizes the propagation of the RF signal inside the dielectric waveguide from the FP to the corresponding PA in the segment. As the signal propagates through the waveguide, it experiences both attenuation and phase variation. The attenuation is mainly caused by dielectric absorption and conductor losses within the waveguide and is characterized by the attenuation coefficient ${\kappa}$ measured in dB/m \cite{P14}. In addition to attenuation, the signal also experiences phase variation while propagating inside the waveguide. This effect is characterized by an effective refractive index, $\eta_{\mathrm{eff}}$, which determines the guided wavelength $\lambda_G=\frac{\lambda}{\eta_{\mathrm{eff}}}$, where  $\lambda=\frac{c}{f_c}$ is the free-space wavelength, c is the speed of light, and $f_c$ is the carrier frequency.  Therefore, the forward in-waveguide channel between the FP and the $m^{th}$ PA of the  $A_l$ segment  is given by 
\vspace{-0.25em}
\begin{equation}
h_{l,m}
=\underbrace{10^{-\frac{\kappa\left\rVert\boldsymbol{\Omega}_{F_l}
-\boldsymbol{\Omega}_{P_{l,m}}\right\rVert}{20}}}_{\text{Attenuation factor}}\;\underbrace{e^{-j\frac{2\pi}{\lambda}\left(\eta_{\mathrm{eff}}
\left\rVert\boldsymbol{\Omega}_{F_l}-\boldsymbol{\Omega}_{P_{l,m}}
\right\rVert\right)}}_{\text{Phase shift}}.
\label{eq:eq1}
\end{equation}
Since there are $M$ PAs in each segment, the signal experiences different attenuation factors and propagation phases before reaching the PA due to different distances from the FP. Thus, the in-waveguide channel vector between FP and different PAs is given as ${\textbf{h}}_{l}=\left[h_{l,1},\,h_{l,2}\,\ldots,\,h_{l,M}\right]\in\mathbb{C}^{1\times M}$. The above channel vector is common for both communication and sensing, but the sensing operation involves an additional reverse echo propagation path. For this one PA in the segment, denoted by $\hat{P}_{l,m}$ is used as the receiving PA that is nearest to the ST. The reflected echo then propagates back to the AP through the dielectric waveguide. Assuming channel reciprocity, the in-waveguide channel environment is the same in both the forward and backward directions and is given by \vspace{-0.25em}
\begin{equation}
\widetilde{h}_{l,m}=10^{-\frac{\kappa \left\rVert\boldsymbol{\Omega}_{\hat{P}_{l,m}}-{\boldsymbol\Omega_{F_l}}\right\rVert}{20}}e^{-j\frac{2\pi}{\lambda}\left(\eta_{\mathrm{eff}} \left\rVert\boldsymbol{\Omega}_{\hat{P}_{l,m}}-{\boldsymbol\Omega_{F_l}}\right\rVert\right)}.
\label{eq:eq3}
\end{equation}

\subsubsection{\textbf{Wireless Channel}}
Although users and targets operate in different propagation environments and experience different channel characteristics, they share the same propagation model. To characterize the LoS condition between the $m^{th}$ PA of $A_q$  and the $k^{th}$ ground user of the $A_1$ segment. A binary random variable $\theta_{q,m,l,k}\in\{0,1\}$ is defined, where
$\theta_{q,m,l,k}=1$ represents the LoS path, for which the Rician distribution is used to model the small-scale fading, whereas $\theta_{q,m,l,k}=0$ represents the NLoS path, for which the Rayleigh distribution is used \cite{A16}. Following the LoS blockage model for ultra-dense indoor environments, the LoS probability is given by \cite{P18}
\begin{equation}
P_{q,m,l,k}=\Pr(\theta_{q,m,l,k}=1)
=e^{-\phi\left\lVert
\boldsymbol{\Omega}_{P_{q,m}}-
\boldsymbol{\Omega}_{\mathrm{CU}_{l,k}}
\right\rVert},
\end{equation}
where $\phi$ denotes the LoS blockage parameter. Accordingly, the probability of the NLoS channel is $1-P_{q,m,l,k}$. Following the LoS blockage model in \cite{P18} and the 
PA-to-user channel model in\cite{A17}, the channel between 
the  $m^{th}$ PA of the  $A_q$ and $k^{th}$ user  of the $A_1$ segment is represented as ${ h}^{q,m}_{{l,k}}$ and given by \vspace{-0.5em}
\begin{equation}
{h}^{q,m}_{l,k}
=
\begin{cases}
\frac{\xi e^{-j\frac{2\pi}{\lambda}
\left\lVert
\boldsymbol{\Omega}_{P_{q,m}}
-\boldsymbol{\Omega}_{AU_{l,k}}
\right\rVert} }
{\left\lVert
\boldsymbol{\Omega}_{P_{q,m}}
-
\boldsymbol{\Omega}_{AU_{l,k}}
\right\rVert},
&\text{$l\neq1$},
\\[2ex]
\frac{\xi G_{q,m,l,k} }
{\left\lVert
\boldsymbol{\Omega}_{P_{q,m}}
-
\boldsymbol{\Omega}_{CU_{l,k}}
\right\rVert^{\nu/2}},
&\text{$l=1$},
\end{cases}
\label{eq:three_channel_models}
\end{equation}
where $\xi=\frac{\lambda}{4\pi}$ is the free-space path-loss at a reference distance of 1 m, $\nu$ is the path-loss exponent, and $G_{q,m,l,k}$ is small-scale fading, and it is define as
\vspace{-0.25em}
\begin{equation}
\begin{split}
\!\!&G_{q,m,l,k}\!=\!
\begin{cases}
\sqrt{\frac{K_R}{K_R+1}}\,
e^{-j\frac{2\pi}{\lambda}
\left\lVert
\boldsymbol{\Omega}_{P_{q,m}}
-\boldsymbol{\Omega}_{CU_{l,k}}
\right\rVert}\\+\sqrt{\frac{1}{K_R+1}}\,g_{q,m,l,k},
& \text{$\theta_{q,m,l,k}=1$},
\\
g_{q,m,l,k},&\text{$\theta_{q,m,l,k}=0$},
\end{cases}
\label{eq:fading_coefficient_G}
\end{split}
\end{equation}
where $K_R$ denotes the Rician factor, and $g_{q,m,l,k}\sim\mathcal{CN}(0,1)$.
\par
At the $k^{th}$ user, the signal is coming from $M$ PAs of $A_q$, so the channel vector  between these  PAs and the $k^ {th}$ user is as $\mathbf{h}^{q}_{{l,k}}=
\begin{bmatrix}h^{q,1}_{{l,k}} &\cdots &h^{q,M}_{{l,k}}
\end{bmatrix}^{\mathsf T}\in\mathbb{C}^{M\times 1}$. The effective channel between the FP and the $k^{th}$ user is given as
\begin{equation}
h_{l,k}^{q,\mathrm{eff}}
=\mathbf{h}_{q}\mathbf{h}^{q}_{l,k}
=\sum\nolimits_{m=1}^{M}h_{q,m}\,h^{q,m}_{l,k}.
\label{eq:effective_channel}
\end{equation}
When the signal comes from the same serving segment, then in that case $q=l$, and \eqref{eq:effective_channel} becomes $h_{l,k}^{l,\mathrm{eff}}=\sum_{m=1}^{M}h_{l,m}\,h^{l,m}_{l,k}$.
\par
Consider the sensing target is in direct LoS with the PA, so the signal radiated by the PA propagates through free space towards the ST. The signal is then reflected by the ST and returns to the selected receiving PA. Therefore, the sensing process consists of two propagation paths: a forward path from the PA to the ST and a return path from the ST to the PA. Accordingly, the forward wireless channel between $m^{th}$ PA of $A_q$  and the ST of the $A_l$ segment is given by   \vspace{-0.5em}
\begin{equation}
h^{q,m}_{{l,T}}=\frac{\xi e^{-j\frac{2\pi}{\lambda}
\left\rVert\boldsymbol{\Omega}_{P_{q,m}}-\boldsymbol{\Omega}_{T_l}\right\rVert} }{\left\rVert
\boldsymbol{\Omega}_{P_{q,m}}-\boldsymbol{\Omega}_{T_l}\right\rVert}
\label{eq:eq9}
\end{equation}
 So, the channel vector between $M$  PAs of $A_q$ and the ST of $A_l$, the segment is as $\mathbf{h}^{q}_{{l,T}}=
\begin{bmatrix}h^{q,1}_{{l,T}} & \cdots & h^{q,M}_{{l,T}}
\end{bmatrix}^{\mathsf T}\in\mathbb{C}^{M\times 1}$. The effective forward sensing channel between the FP of $A_q$ and the ST of $A_l$ is expressed as \vspace{-0.5em}
\begin{equation}
\check{h}_{l,\mathrm{sen}}^{q}
=\mathbf{h}_{q}\mathbf{h}^{q}_{l,T}
=\sum\nolimits_{m=1}^{M}h_{q,m}\,h^{q,m}_{l,T}.
\label{eq:effective_sensing_channel}
\end{equation}
Like communication, when the signal comes from the same segment, then \eqref{eq:effective_sensing_channel} becomes $\check{h}_{l,\mathrm{sen}}=\sum_{m=1}^{M}h_{l,m}\,h^{l,m}_{l,T_l}$. Upon reaching the ST, the received signal spreads isotropically. Therefore, only a small fraction of the signal is reflected back towards the receiving PA, which is characterised by the radar cross section (RCS), denoted by $\sigma_T$. The RCS represents the effective scattering area of the ST and determines the strength of the reflected radar signal. The PA closest to the ST is selected as the receiving PA denoted as $\hat{P}_{l,m}$, and the sensing receiver is assumed to be co-located with it, following the monostatic PASS-assisted ISAC sensing framework. Therefore, the reflected return path is modelled through the ST to the $\hat{P}_{l,m}$ path, followed by receiver-side in-waveguide propagation. The selected receiver PA, $\hat{P}_{l,m}$ is chosen by the nearest section rule and expressed as $\hat{P}_{l,m}
=P_l{\arg\min_{m\in\{1,\ldots,M\}}\,\left\|\boldsymbol{\Omega}_{T_l}
-\boldsymbol{\Omega}_{P_{l,m}}\right\|}$ and the coordinates of the selected PA can be given as $ \boldsymbol{\Omega}_{\hat{P}_{l,m}}$, where $m=\boldsymbol{\Omega}_{P_l{\arg\min_{m\in\{1,\ldots,M\}}\,\left\|\boldsymbol{\Omega}_{T_l}
-\boldsymbol{\Omega}_{P_{l,m}}\right\|}}$ \cite{S3}. Since ST is aerial and has direct visibility with the PA, we modelled it as a direct LoS link. The reflected echo then propagates from the ST to the selected receiving PA through the return wireless channel, which is given by  \vspace{-1em}
\begin{equation}
\widetilde{h}_{T,l}
=
\frac{
\xi\sqrt{\dfrac{\sigma_T}{4\pi}}e^{-j\frac{2\pi}{\lambda}\left\rVert\boldsymbol{\Omega}_{T_l}-\boldsymbol{\Omega}_{\hat{P}_{l,m}}
\right\rVert}
}{\left\rVert\boldsymbol{\Omega}_{T_l}-\boldsymbol{\Omega}_{\hat{P}_{l,m}}
\right\rVert}.
\label{eq:eq11}
\end{equation}
where, $\sqrt{\sigma_T/(4\pi)}$ represents the target-scattering amplitude associated with the RCS $\sigma_T$. After reflection from the ST, the echoes are received by the selected sensing PA and guided back to the AP through the dielectric waveguides for further signal processing. By using \eqref{eq:eq3} and \eqref{eq:eq11}, the effective backward sensing channel between the ST and FP is $ \widetilde{h}_{l,\mathrm{sen}}
=\widetilde{h}_{T,l}\,\widetilde{h}_{l,m}$ and the overall sensing channel is expressed as  \vspace{-0.5em}
\begin{equation}
h_{l,\mathrm{sen}}
=
\check{h}_{l,\mathrm{sen}}
\widetilde{h}_{l,\mathrm{sen}}
=
\Big(
\sum\nolimits_{m=1}^{M}
h_{l,m}\,h^{l,m}_{l,T_l}
\Big)
\widetilde{h}_{T,l}\,
\widetilde{h}_{l,m}.
\label{eq:overall_sensing_channel}
\end{equation}
\vspace{-2.5em}
\subsection{Signal Model}
Let the total transmit power $\mathcal{P}_t$ be equally shared among the $L$ segments. Also, in this work, we assume that the transmitted power is equally distributed among the $M$ PAs, so the power of each PA is denoted by $\mathcal{P}_{l_{m}}$ and given by $\mathcal{P}_{l_{m}}=\frac{\mathcal{P}_t}{LM}$. Since each segment performs both communication and sensing simultaneously, its allocated power is further divided between these two functionalities. The communication power is used to serve the $K$ users in the segment using non-orthogonal multiple access (NOMA), whereas the remaining sensing power is used to transmit sensing signals toward the ST. Accordingly, the combined ISAC signal transmitted by the $A_l$ segment is expressed as \vspace{-0.75em}
\begin{equation}
\psi_l(t)=\underbrace{\sum\nolimits_{i=1}^{K}\sqrt{b_{l,i}\alpha}\,x_{l,i}(t)}_{\text{NOMA communication signal}}
+\underbrace{\sqrt{(1-\alpha)}\,s_l(t)}_{\text{Sensing probing signal}},
\label{eq:eq12}
\end{equation}
where $b_{l,i}$ is the NOMA power allocation coefficient corresponding to the information symbol $x_{l,i}$ intended for the communication user such that $\sum_{i=1}^{K} b_{l,i}=1$ and ${E}\left[|x_{l,i}|^2\right]=1$, while $s_l$  denotes the sensing signal associated with ST with ${E}\left[|s_l|^2\right]=1$. We also assume that the sensing waveform of the $A_l$ segment is orthogonal to the sensing waveforms of adjacent segments, $\langle s_l,s_q\rangle=0$, $\forall\, q\neq l$. $\alpha$ and $1-\alpha$ represent the fractions of the transmit power allocated to communication and sensing. Statistical channel state information (CSI) is assumed to be available at the AP for user ordering and power allocation, while each user is assumed to have perfect knowledge of its own effective channel for signal detection. We assume that ${E}\left[\left|h_{l,1}^{l,\mathrm{eff}}\right|^{2}\right]>\cdots>{E}\left[\left|h_{l,K}^{l,\mathrm{eff}}\right|^{2}\right]$, so the NOMA power-allocation coefficients of the user are $ 0< b_{l,1}<\cdots<b_{l,K}<1$ \cite{S4}. Based on this signal model, in the next section, we discuss the received signal and SINR calculation in detail.
Without loss of generality, we consider only the interference from the adjacent segment, as the remaining segments are sufficiently far from the considered segment. Consequently, their interference is assumed to be negligible. For a vertically segmented system, $\mathcal{I}_l$  denotes the set of segments adjacent to the $A_l$  and defined as \cite{S7,A4}
\begin{equation}
\mathcal{I}_l=
\begin{cases}
\{\emptyset\},& L=1,\\
\{2\},& L>1, l=1,\\
\{l-1,l+1\},& L>1, 1<l<L,\\
\{L-1\},& L>1, l=L.
\end{cases}
\label{eq:eq15}
\end{equation}
\subsubsection{{\textbf{Communication functionality }}}
The superimposed ISAC signal received by the $k^{th}$ user associated with the $l^{th}$ segment  is given by \vspace{-0.5em}
\begin{equation}
\begin{split}
&y_{l,k}(t)
\\&=
\begin{cases}
\sqrt{\mathcal{P}_{1_{m}}}\Big(
h_{1,k}^{1,\mathrm{eff}}\psi_1(t)
+h_{1,k}^{2,\mathrm{eff}}\psi_{2}(t)
\Big)+ n_{1,k}, & \hspace{-15mm}l=1,
\\[1mm]
\sqrt{\mathcal{P}_{l_{m}}}\Big(
h_{l,k}^{{l-1},\mathrm{eff},}\psi_{l-1}(t)+
h_{l,k}^{l,\mathrm{eff}}\psi_l(t)\\
+h_{l,k}^{{l+1},\mathrm{eff},}\psi_{l+1}(t)
\Big)+n_{l,k}, &\hspace{-20mm} 1<l<L,
\\[1mm]
\sqrt{\mathcal{P}_{l_{m}}}\Big(
h_{L,k}^{{L-1},\mathrm{eff}}\psi_{L-1}(t)
+h_{L,k}^{L,\mathrm{eff}}\psi_L(t)
\Big)+n_{L,k},\hspace{1mm} l=L,
\\[1mm]
\end{cases}
\label{eq:eq16}
\end{split}
\end{equation}
where $n_{l,k}\sim\mathcal{CN}(0,\sigma_{l,k}^2)$ denotes the additive white Gaussian noise (AWGN) with noise variance $\sigma_{l,k}^2$. After applying SIC at the receiver, the SINR of the $k^{th}$ user is expressed as \vspace{-0.25em}
\begin{equation}
\gamma_{l,k}
=\frac{
\mathcal{P}_{l_{m}}\left\lvert h_{l,k}^{l,\mathrm{eff}}\right\rvert^2
\alpha b_{l,k}}{I_{intra}+\sum_{q\in\mathcal{I}_l}\mathcal{P}_{l_{m}}\left\lvert h_{l,k}^{q,\mathrm{eff}}\right\rvert^2+\sigma_{l,k}^2},
\label{eq:eq17}
\end{equation}
where $I_{intra}= \mathcal{P}_{l_{m}}\left\lvert h_{l,k}^{l,\mathrm{eff}}\right\rvert^2
\left(\alpha\sum_{u=1}^{k-1}b_{l,u}+(1-\alpha)\right)$.
\vspace{.2cm}
\paragraph{{ \textbf{Ground users}}}
By putting $l=1$ in \eqref{eq:eq16}, and after SIC is applied at the receiver, the SINR of the $k^{th}$ user of the $A_1$ segment  is expressed as \vspace{-0.5em}
\begin{equation}
\gamma_{1,k}
=\frac{\mathcal{P}_{l_{m}}\left\lvert h_{1,k}^{1,\mathrm{eff}}\right\rvert^2
\alpha b_{1,k}}{I_{intra}+\mathcal{P}_{l_{m}}\left\lvert h_{1,k}^{2,\mathrm{eff}}\right\rvert^2+\sigma_{1,k}^2}.
\label{eq:eq18}
\end{equation}
Using equation \eqref{eq:eq18}, the rate is given by $\mathcal{R}_{1,k}=\log_2(1+\gamma_{1,k})$.
\paragraph{{\textbf{Aerial users}}} 
 Without loss of generality, we consider $l=2$ for the aerial case, for which the adjacent segments are $A_1,A_3$, so by putting $l=2$, in \eqref{eq:eq16}. After performing SIC at the receiver, the SINR of the $k^{th}$ user of $A_2$ is expressed as 
\begin{equation}
\gamma_{2,k}
=\frac{
\mathcal{P}_{l_m}\left\lvert h_{2,k}^{2,\mathrm{eff}}\right\rvert^2
\alpha b_{2,k}}{I_{intra}+\mathcal{P}_{l_{m}}\left\lvert h_{2,k}^{1,\mathrm{eff}}\right\rvert^2+\mathcal{P}_{l_{m}}\left\lvert h_{2,k}^{3,\mathrm{eff}}\right\rvert^2+\sigma_{2,k}^2}.
\label{eq:eq19}
\end{equation}
Using equation \eqref{eq:eq19}, the rate is given as $\mathcal{R}_{2,k}=\log_2(1+\gamma_{2,k})$
\paragraph{{\textbf{Segment sum rate calculation}}}
The segment rate of the users in the segment $A_l$  is given as $\mathcal{R}_{l}^{\mathrm{sum}}=\sum_{k=1}^K\mathcal{R}_{l,k}$, where $\mathcal{R}_{l,k}$ denotes the instantaneous rate of the $k^{th}$ user in $A_l$.
\subsubsection{\textbf{Sensing functionality}} After propagating through the wireless channel, the transmitted signal is reflected by the target and received back at the sensing receiver, which is used for target range estimation. Accordingly, the received sensing signal is expressed as
\begin{multline}
y_{l}^{\rm sen}(t)=
\sqrt{\mathcal{P}_{l_m}}\Big(\check{h}_{l-1,\rm sen}\widetilde{h}_{l,\rm sen}\psi_{l-1}(t-\tau)+h_{l,\rm sen}\psi_l(t-\tau)\\+ \check{h}_{l+1,\rm sen}
 \widetilde{h}_{l,\rm sen} \psi_{l+1}(t-\tau)\Big) + n_s,
\label{eq:eq20}
\end{multline}
where $\tau$ the round trip propagation delay, and  $n_{s}\sim\mathcal{CN}(0,\sigma_{l,s}^2)$ denotes the AWGN  sensing noise. Practically, the signals radiated from different PAs experience different round-trip propagation delays due to the differences in their propagation paths and distances. To represent these different delays by a common round-trip delay $\tau$, we adopt the common delay assumption as in \cite{S3,S6}. This assumption is valid when the maximum PA deployment length of a segment satisfies $A_{l,\mathrm{seg}}\leq \frac{c}{B}$, where $B$ denotes the bandwidth.  Since the receiving PA is co-located with the transmitting PA, the received sensing signal experiences self-communication interference. It is assumed that the sensing receiver employs perfect self-interference cancellation, such as analog, digital, or hybrid cancellation techniques, to completely remove this self-communication interference. In addition, the reflected communication component of the  ISAC signal interferes with the desired sensing signal of the $A_l$ segment. However, since the transmitted communication signal is known at the AP, its reflected component is reconstructed and cancelled at the sensing receiver. Moreover, the sensing waveforms of adjacent segments are assumed to be mutually orthogonal, such that the sensing interference from adjacent segments is eliminated after matched filtering. After removing all this interference, the SINR of the received sensing signal is given by 
\begin{equation}
\gamma_{l}^{\rm sen}
=
\frac{
\mathcal{P}_{l_m}(1-\alpha)
\left\lvert h_{l,\rm sen}\right\rvert^2
}{
\mathcal{P}_{l_m}\alpha
\sum_{q\in\mathcal{I}_l}
\left\lvert
\check{h}_{l,\rm sen}^{q}
\widetilde{h}_{l,\rm sen}
\right\rvert^2
+
\sigma_{l,s}^{2}
}.
\label{eq:eq22}
\end{equation}

\vspace{-0.5em}
 \section{Performance Analysis}
 \label{sec: performance analysis}
In this section, the analytical performance of the proposed V-SWAN-assisted downlink NOMA-ISAC EDMA framework is analysed for a generalised multi-user and multi-PA scenario. The effective channels are formed by the superposition of multiple cascaded in-waveguide and wireless propagation channels. Due to the combined effects of in-waveguide attenuation, phase variations, and varying wireless propagation distances across different PAs, the resulting composite channel exhibits a highly complicated statistical distribution. Therefore, obtaining an exact closed-form expression is analytically intractable. 
\vspace{-0.5em}
\subsection{\textbf{Channel characterization of users} }
 The outage analysis in multi-PA and multi-user systems requires a statistical characterization of channels to evaluate the performance of the proposed method for multi-user systems. During the analysis, the PAs are uniformly distributed within the segment and the effective channel is obtained by summing the contribution from multiple PAs\footnote{For mathematical tractability, the channels from each PA to user is considered to be independent. In practical scenarios, a minimum inter-PA spacing of $\frac{\lambda}{2}$ is usually imposed to mitigate coupling. However, for the considered system parameters, imposing this only removes a limited subset of PA configurations and does not materially alter the channel statistics. Hence, its impact on the performance is negligible as discussed in \cite{S3}.} as shown in \eqref{eq:effective_channel}. Such type of channel distribution along with summation is consistent with other PASS related works \cite{P4,S3}. Now, the channel between the FP and a user consists of two cascaded channels, i.e. the in-waveguide channel between the FP and the PA and the wireless channel between the PA and the corresponding user. Since both the PA locations and the user locations are randomly distributed, the effective channel is governed by their joint spatial distribution. Accordingly, the statistical characterization is performed first followed by the respective outage analysis. 
 \subsubsection{\textbf{Ground Users}}
 The statistical characterization of the CU effective channel is performed by conditioning on CU and PA positions. So, the randomness comes from the LoS/NLoS state and the associated small-scale fading. Therefore, the conditioned effective channel is non-zero mean because the channel is a mixture of Rayleigh and Rician distributions. A moment-matching approach is used to approximate the channel as the effective channel as a complex Gaussian random variable. For this, the effective mean and variance are required. From \eqref{eq:three_channel_models} and \eqref{eq:fading_coefficient_G}, the conditional  mean and variance of the LoS/NLoS channel are given by
 \vspace{-0.0em}
\begin{equation}
\begin{split}
&\mu_{l,k}^{q,m}\left(\boldsymbol{\Omega}_{P_{q,m}},\boldsymbol{\Omega}_{CU_{l,k}}\right)\\&\hspace{10mm}=
\begin{cases}
\frac{\xi\sqrt{\frac{K_R}{K_R+1}}e^{-j\frac{2\pi}{\lambda}
\left\lVert
\boldsymbol{\Omega}_{P_{q,m}}
-
\boldsymbol{\Omega}_{CU_{l,k}}
\right\rVert}}
{\left\lVert
\boldsymbol{\Omega}_{P_{q,m}}
-
\boldsymbol{\Omega}_{CU_{l,k}}
\right\rVert^{\frac{\nu}{2}}},
& \hspace{-.2cm}\text{$\theta_{q,m,l,k}=1$},\\
$0$, & \hspace{-.2cm}\text{$\theta_{q,m,l,k}=0$},
\label{eq:los_mean}
\end{cases}
\end{split}
\end{equation}
and  
\begin{equation}
\begin{split}
&\left(\sigma_{l,k}^{{q,m}}\left(
\boldsymbol{\Omega}_{P_{q,m}},
\boldsymbol{\Omega}_{CU_{l,k}}
\right)\right)^2
\\&\hspace{15mm}=
\begin{cases}
\frac{\xi^2}{\left\lVert\boldsymbol{\Omega}_{P_{q,m}}
-\boldsymbol{\Omega}_{CU_{l,k}}\right\rVert^{\nu} \left(K_R+1\right)},
& \hspace{-.2cm}\text{$\theta_{q,m,l,k}=1$},\\
\frac{\xi^2}{\left\lVert\boldsymbol{\Omega}_{P_{q,m}}-\boldsymbol{\Omega}_{CU_{l,k}}\right\rVert^{\nu}},
& \hspace{-.2cm}\text{$\theta_{q,m,l,k}=0$}.\\
\label{eq:los_varince}
\end{cases}
\end{split}
\end{equation}
 By using \eqref{eq:los_mean} and the law of total expectation \cite{l1}, the mean of the composite single-link channel is obtained as
\begin{align}
&\mu_{l,k}^{q,m}
\left(
\boldsymbol{\Omega}_{P_{q,m}},
\boldsymbol{\Omega}_{CU_{l,k}}
\right)\nonumber\\
&\hspace{1.5em}=
P_{q,m,l,k}
\left(
\mu_{l,k}^{q,m}
\left(
\boldsymbol{\Omega}_{P_{q,m}},
\boldsymbol{\Omega}_{CU_{l,k}}
\right)
\mid
\theta_{q,m,l,k}=1
\right).
\end{align}
 By using \eqref{eq:los_varince} and the law of total variance \cite{l2}, the variance of the composite single-link channel is given as in \eqref{eq:variance}.
 \begin{figure*}[!t]
  \begin{equation}
\begin{split}
&\left(\sigma_{l,k}^{q,m}\left(\boldsymbol{\Omega}_{P_{q,m}},\boldsymbol{\Omega}_{CU_{l,k}}\right)\right)^2\\&=
P_{q,m,l,k}
\left(
\sigma_{l,k}^{q,m}
\left(
\boldsymbol{\Omega}_{P_{q,m}},
\boldsymbol{\Omega}_{CU_{l,k}}
\right)
\mid
\theta_{q,m,l,k}=1
\right)^2+
\left(1-P_{q,m,l,k}\right)\times\left(
\sigma_{l,k}^{q,m}
\left(
\boldsymbol{\Omega}_{P_{q,m}},
\boldsymbol{\Omega}_{CU_{l,k}}
\right)
\mid
\theta_{q,m,l,k}=0
\right)^2
\\&
\hspace{80mm}+P_{q,m,l,k}\left(1-P_{q,m,l,k}\right)
\left|
\mu_{l,k}^{q,m}
\left(
\boldsymbol{\Omega}_{P_{q,m}},
\boldsymbol{\Omega}_{CU_{l,k}}
\right)
\mid
\theta_{q,m,l,k}=1
\right|^2 .
\end{split}
\label{eq:variance}
\end{equation}
 \vspace{-1.25em}
\end{figure*}
Let $z$ denote the generic position of the $m^{th}$ PA along the $A_q$ segment, and its position along the waveguide is $\boldsymbol{\Omega}_{P_{q,m}}=[0,0,z]$. Assuming statistically independent and identically distributed PAs. Therefore, the effective channel can be approximated as $h_{l,k}^{q,\mathrm{eff}}
\sim\mathcal{CN}
\left(\mu_{l,k}^{\mathrm{q,eff}}\left(\boldsymbol{\Omega}_{CU_{l,k}}\right),\left(\sigma_{l,k}^{q,\mathrm{eff}}\left(\boldsymbol{\Omega}_{CU_{l,k}}\right)\right)^2\right)$, where the mean and variance of $h_{l,k}^{q,\mathrm{eff}}$ are given by
\begin{equation}
\begin{split}
&{\mu}_{l,k}^{q,\mathrm{eff}}
\left(\boldsymbol{\Omega}_{CU_{l,k}}\right)
\\&\hspace{8mm}=\frac{M}{A_{q,\mathrm{seg}}}
\int_{A_{q,\mathrm{seg,min}}}^{A_{q,\mathrm{seg,max}}}
h_{q,m}(z)\mu_{l,k}^{q,m}
\left(z,\boldsymbol{\Omega}_{CU_{l,k}}\right)
\,\mathrm{d}z,
\end{split}
\label{eq:cu_mean_pa}
\end{equation}
and  \eqref{eq:effective_variance},
where $A_{q,seg,\max}=q{A_{q,seg}}$ and $A_{q,seg,\min}=(q-1){A_{q,seg}}$.
\begin{figure*}[!t]
\begin{equation}
\begin{split}
\left({\sigma}_{l,k}^{q,\mathrm{eff}}
\left(\boldsymbol{\Omega}_{CU_{l,k}}\right)\right)^2=
\frac{M}{A_{q,\mathrm{seg}}}
\int_{A_{q,\mathrm{seg,min}}}^{A_{q,\mathrm{seg,max}}}
|h_{q,m}(z)|^2
\left(
\sigma_{l,k}^{q,m}
\left(z,\boldsymbol{\Omega}_{CU_{l,k}}\right)
\right)^2
\,\mathrm{d}z
 + \\&\hspace{-45mm}
\frac{M}{A_{q,\mathrm{seg}}}
\int_{A_{q,\mathrm{seg,min}}}^{A_{q,\mathrm{seg,max}}}
\left|
h_{q,m}(z)
\mu_{l,k}^{q,m}
\left(z,\boldsymbol{\Omega}_{CU_{l,k}}\right)
\right|^2
\,\mathrm{d}z- 
\frac{1}{M}\left|{\mu}_{l,k}^{q,\mathrm{eff}}
\left(\boldsymbol{\Omega}_{CU_{l,k}}\right)
\right|^2 .
\end{split}
\label{eq:effective_variance}
\end{equation}
\hrule
\vspace{-1em}
\end{figure*}
Accordingly, the effective channel power is defined as
$H_{l,k}^{\mathrm{q,eff}}=\left|h_{l,k}^{\mathrm{q,eff}}\right|^2$. Since the effective channel has non-zero mean, the normalized channel power follows a scaled noncentral chi-square distribution with two degrees of freedom. The PDF of the effective channel power is given by
\begin{equation}
\begin{split}
&f_{H_{l,k}^{\mathrm{q,eff}}}
(x\mid\boldsymbol{\Omega}_{CU_{l,k}})\\&\hspace{11mm}=
\frac{e^{-\frac{x+\left|\mu_{l,k}^{\mathrm{q,eff}}\left(\boldsymbol{\Omega}_{CU_{l,k}}\right)\right|^2}{
\left(\sigma_{l,k}^{ q,\mathrm{eff}}\left(\boldsymbol{\Omega}_{CU_{l,k}}\right)\right)^2}}I_0\left(\frac{2\left|\mu_{l,k}^{\mathrm{q,eff}}\left(\boldsymbol{\Omega}_{CU_{l,k}}\right)\right|\sqrt{x}}{\left(\sigma_{l,k}^{q,\mathrm{eff}}\left(\boldsymbol{\Omega}_{CU_{l,k}}\right)\right)^2}\right)}{\left(\sigma_{l,k}^{q,\mathrm{eff}}\left(\boldsymbol{\Omega}_{CU_{l,k}}\right)\right)^2},
\end{split}
\label{eq:effective_power_pdf}
\end{equation}
where $I_0(\cdot)$ denotes the modified Bessel function of the first kind and zeroth order. Accordingly, the closed form of the CDF of the effective channel power is given by the Marcum $\mathrm{Q}_1$ function as 
\begin{equation}
\begin{split}
&F_{H_{l,k}^{\mathrm{q,eff}}}(x\mid\boldsymbol{\Omega}_{CU_{l,k}})
\\&\hspace{4mm}=1-\mathrm{Q}_1\left(\frac{\sqrt{2}\left|\mu_{l,k}^{\mathrm{q,eff}}
\left(\boldsymbol{\Omega}_{CU_{l,k}}\right)\right|}{\sigma_{l,k}^{q,\mathrm{eff}}\left(\boldsymbol{\Omega}_{CU_{l,k}}\right)},
\frac{\sqrt{2x}}{
\sigma_{l,k}^{\mathrm{q,eff}}\left(\boldsymbol{\Omega}_{CU_{l,k}}\right)}\right).
\label{eq:effective_power_cdf}
\end{split}
\end{equation}
\subsubsection{\textbf{Aerial Users }}
The statistical characterization of the AU effective channel is performed by conditioning on AU positions. The randomness comes due to the random distribution of PA. Using the moment-matching approximation, the effective AU channel is approximated as $ h_{l,k}^{q,\mathrm{eff}}\sim\mathcal{CN}\left(\mu_{l,k}^{q,\mathrm{eff}}\left(\boldsymbol{\Omega}_{AU_{l,k}}\right),\left(\sigma_{l,k}^{q,\mathrm{eff}}\left(\boldsymbol{\Omega}_{AU_{l,k}}\right)\right)^2\right)$. Let $z$ denote the generic position of the $m^{\mathrm{th}}$ PA along the $A_q$ segment and its position along the waveguide is $\boldsymbol{\Omega}_{P_{q,m}}=[0,0,z]$. The mean and second moment of a single PA contribution, after averaging over the PA position, are respectively given by
\begin{equation}
\mu_{l,k}^{q,m}\left(\boldsymbol{\Omega}_{AU_{l,k}}\right)
=\frac{1}{A_{q,seg}}
\int_{A_{q,seg,\min}}^{A_{q,seg,\max}}
h_{q,m}(z)h_{l,k}^{q,m}(z)
\,\mathrm{d}z,
\label{eq:au_mean_pa}
\end{equation}
and
\begin{equation}
\Psi_{l,k}^{q,m}\left(\boldsymbol{\Omega}_{AU_{l,k}}\right)
=\frac{1}{A_{q,seg}}
\int_{A_{q,seg,\min}}^{A_{q,seg,\max}}
\left|
h_{q,m}(z)h_{l,k}^{q,m}(z)
\right|^2
\,\mathrm{d}z.
\label{eq:au_second_moment}
\end{equation}
Hence, the variance of a single PA contribution can be given as  $\left(\sigma_{l,k}^{q,m}\left(\boldsymbol{\Omega}_{AU_{l,k}}\right)\right)^2=\Psi_{l,k}^{q,m}-$$\left|\mu_{l,k}^{q,m}\right|^2$. Assuming statistically independent and identically distributed PA contributions, the mean and variance of the effective channel are $ \mu_{l,k}^{\mathrm{q,eff}}\left(\boldsymbol{\Omega}_{AU_{l,k}}\right)= M\mu_{l,k}^{q,m}\left(\boldsymbol{\Omega}_{AU_{l,k}}\right)$ and $ \left(\sigma_{l,k}^{q,\mathrm{eff}}\left(\boldsymbol{\Omega}_{AU_{l,k}}\right)\right)^2=M\left(\sigma_{l,k}^{q,m}\left(\boldsymbol{\Omega}_{AU_{l,k}}\right)\right)^2$. The effective AU channel power gain as $ H_{l,k}^{q,\mathrm{eff}} =\left|h_{l,k}^{q,\mathrm{eff}}\right|^2$. Since the effective channel has a non-zero mean, the normalized channel power follows a noncentral chi-square distribution with two degrees of freedom. The  PDF and CDF of the effective channel are given as  \vspace{-0.5em}
\begin{equation}
\begin{split}    
&f_{H_{l,k}^{\mathrm{q,eff}}}(x\mid\boldsymbol{\Omega}_{AU_{l,k}})
\\&\hspace{11mm}=\frac{e^{-\frac{x+\left|\mu_{l,k}^{q,\mathrm{eff}}\left(\boldsymbol{\Omega}_{AU_{l,k}}\right)\right|^2}{\left(\sigma_{l,k}^{q,\mathrm{eff}}\left(\boldsymbol{\Omega}_{AU_{l,k}}\right)\right)^2}}I_0\left(\frac{2\left|\mu_{l,k}^{q,\mathrm{eff}}\left(\boldsymbol{\Omega}_{AU_{l,k}}\right)\right|\sqrt{x}}{\left(\sigma_{l,k}^{q,\mathrm{eff}}\left(\boldsymbol{\Omega}_{AU_{l,k}}\right)\right)^2}\right)}{\left(\sigma_{l,k}^{q,\mathrm{eff}}\left(\boldsymbol{\Omega}_{AU_{l,k}}\right)\right)^2
},
\label{eq:au_pdf}
\end{split}
\end{equation}
\vspace{-0.5em}
\begin{equation}
\begin{split}
&F_{H_{l,k}^{q,\mathrm{eff}}}(x\mid\boldsymbol{\Omega}_{AU_{l,k}})
\\& \hspace{4mm}=1-\mathrm{Q}_1\left(\frac{\sqrt{2}\left|\mu_{l,k}^{\mathrm{q,eff}}\left(\boldsymbol{\Omega}_{AU_{l,k}}\right)\right|}{\sigma_{l,k}^{q,\mathrm{eff}}\left(\boldsymbol{\Omega}_{AU_{l,k}}\right)},\frac{\sqrt{2x}}{\sigma_{l,k}^{q,\mathrm{eff}}\left(\boldsymbol{\Omega}_{AU_{l,k}}\right)}\right).
\label{eq:au_cdf}
\end{split}
\end{equation}

\vspace{-1em}
\subsection{\textbf{ Channel characterization of Target:}}
Similar to the aerial case, the forward channel of sensing shown in \eqref{eq:effective_sensing_channel} is characterized by conditioning on a fixed sensing target location. Let $z$ denote the generic position of the $m^{\mathrm{th}}$ PA along the $A_q$ segment. By using a moment-matching approach, the forward  effective channel is approximated as $ h_{l,T_l}^{q,\mathrm{eff}}
\sim\mathcal{CN}\left(\mu_{l,T_l}^{\mathrm{q,eff}}\left(\boldsymbol{\Omega}_{T_l}\right),\left(\sigma_{l,T_l}^{\mathrm{q,eff}}\left(\boldsymbol{\Omega}_{T_l}\right)\right)^2\right)$.  
The conditional  mean and second moment for single  the forward channel are  given by
\begin{equation}
\mu_{l,T_l}^{q,m}\left(\boldsymbol{\Omega}_{T_l}\right)
=\frac{1}{{A_{q,seg}}}
\int_{A_{q,seg,\min}}^{A_{q,seg,\max}}
h_{q,m}(z)h_{l,T_l}^{q,m}(z)
\,\mathrm{d}z,
\label{eq:sen_forward_mean_pa}
\end{equation}
\vspace{-0.75em}
\begin{equation}
\Psi_{l,T_l}^{q,m}\left(\boldsymbol{\Omega}_{T_l}\right)
=\frac{1}{{A_{q,seg}}}
\int_{A_{q,seg,\min}}^{A_{q,seg,\max}}
\left|
h_{q,m}(z)h_{l,T_l}^{q,m}(z)
\right|^2
\,\mathrm{d}z.
\label{eq:sen_forward_second_moment}
\end{equation}
Hence, the variance of the  single forward channel PA is given by
\begin{equation}
\left(\sigma_{l,T_l}^{q,m}\left(\boldsymbol{\Omega}_{T_l}\right)\right)^2=\Psi_{l,T_l}^{q,m}\left(\boldsymbol{\Omega}_{T_l}\right)-\left|\mu_{l,T_l}^{q,m}\left(\boldsymbol{\Omega}_{T_l}\right)\right|^2.
\label{eq:sen_forward_var_pa}
\end{equation}
Accordingly, the mean and variance of the effective forward sensing channel are $\mu_{l,T_l}^{\mathrm{q,eff}} \left(\boldsymbol{\Omega}_{T_l}\right)=M\mu_{l,T_l}^{\mathrm{q,m}}\left(\boldsymbol{\Omega}_{T_l}\right)$, and $\left(\sigma_{l,T_l}^{q,\mathrm{eff}}\left(\boldsymbol{\Omega}_{T_l}\right)\right)^2=M\left(\sigma_{l,T_l}^{\mathrm{q,m}}\left(\boldsymbol{\Omega}_{T_l}\right)\right)^2$, the effective forward sensing-channel power gain as $ H_{l,T_l}^{q,\mathrm{eff}}
=\left|h_{l,T_l}^{q,\mathrm{eff}}\right|^2$. The PDF and  CDF are given as \vspace{-0.0em}
\begin{equation}
\begin{split}    
\!\!\!\!f_{H_{l,T_l}^{\mathrm{q,eff}}}(x\mid\boldsymbol{\Omega}_{T_l})=\frac{e^{-\frac{x+\left|\mu_{l,T_l}^{q,\mathrm{eff}}\left(\boldsymbol{\Omega}_{T_l}\right)\right|^2}{\left(\sigma_{l,T_l}^{q,\mathrm{eff}}\left(\boldsymbol{\Omega}_{T_l}\right)\right)^2}}I_0\left(\frac{2\left|\mu_{l,T_l}^{q,\mathrm{eff}}\left(\boldsymbol{\Omega}_{T_l}\right)\right|\sqrt{x}}{\left(\sigma_{l,T_l}^{q\mathrm{eff}}\left(\boldsymbol{\Omega}_{T_l}\right)\right)^2}\right)}{\left(\sigma_{l,T_l}^{q,\mathrm{eff}}\left(\boldsymbol{\Omega}_{T_l}\right)\right)^2},
\end{split}
\label{eq:sen_forward_pdf}
\end{equation}
\vspace{-0.75em}
\begin{equation}
F_{H_{l,T_l}^{q,\mathrm{eff}}}(x\mid\boldsymbol{\Omega}_{T_l})=1-\mathrm{Q}_1\left(\frac{\sqrt{2}\left|\mu_{l,T_l}^{\mathrm{q,eff}}\left(\boldsymbol{\Omega}_{T_l}\right)\right|}{\sigma_{l,T_l}^{\mathrm{q,eff}}\left(\boldsymbol{\Omega}_{T_l}\right)},\frac{\sqrt{2x}}{\sigma_{l,T_l}^{\mathrm{q,eff}}\left(\boldsymbol{\Omega}_{T_l}\right)}\right).
\label{eq:sen_forward_cdf}
\end{equation}
During sensing, all the PAs transmit the sensing signal toward the ST. So, the forward effective channel is formed by all transmitting PAs. For echo reception, only one ${\hat{P}_l}$ is selected as the receiving PA, which is nearest to ST, so the nearest selection rule gives the distribution. Conditioned on $\boldsymbol{\Omega}_{T_l}$ and $\boldsymbol{\Omega}_{\hat{P}_l}$, the return-side channel power gain, represented as
$\widetilde{H}_{l,\mathrm{sen}}
=\left|\widetilde{h}_{l,\mathrm{sen}}\right|^{2}$,
is deterministic. Let the complete sensing-channel power gain be expressed as
$H_{\mathrm{sen}}=H_{l,T_l}^{q,\mathrm{eff}}\widetilde{H}_{l,\mathrm{sen}}$, the conditional CDF of $H_{\mathrm{sen}}$ is given by
\begin{equation}
\begin{split}
&F_{H_{\mathrm{sen}}}
\left(\mathcal{P}_{2_{m}}\left\lvert h_{2,1}^{2,\mathrm{eff}}\right\rvert^2\left( 
\alpha b_{2,1}+1-\alpha \right)
x\mid
\boldsymbol{\Omega}_{T_l},
{\boldsymbol{\Omega}}_{\hat{P}_l}
\right)
\\&=\Pr\left(H_{\mathrm{sen}}\leq x\mid
\boldsymbol{\Omega}_{T_l},{\boldsymbol{\Omega}}_{\hat{P}_l}\right)
\\&=
\Pr\left(H_{l,T_l}^{q,\mathrm{eff}}
\leq\frac{x}{\widetilde{H}_{l,\mathrm{sen}}}\mid
\boldsymbol{\Omega}_{T_l},{\boldsymbol{\Omega}}_{\hat{P}_l}
\right)
\\&= F_{H_{l,T_l}^{q,\mathrm{eff}}}
\left(\frac{x}{\widetilde{H}_{l,\mathrm{sen}}}
\mid\
\boldsymbol{\Omega}_{T_l}
\right).
\end{split}
\end{equation}
For uniformly distributed PAs along the Z axis with segment length $A_{l,seg}$ and minimum vertical distance is defined as $r=\left|{Z}_{\hat{P}_l}-Z_{T_l}\right|$. The conditional PDF of the nearest selected PA can be written as \vspace{-0.75em}
\begin{equation}
f_{{z}_{\hat{P}_l}}
\left({Z}_{\hat{P}_l}\mid Z_{T_l}
\right)\!=\!\frac{M}{A_{l,seg}}
\left(1-\frac{\mathcal{W}
\left({Z}_{\hat{P}_l},Z_{T_l}
\right)}{A_{l,seg}}
\right)^{M-1}\!\!,
\label{eq:nearest_pa_pdf}
\end{equation}
where $\mathcal{W}(Z_{\hat{P}_l},Z_{T_l})
=[\min\{z_{\mathrm{upper},A_l},Z_{T_l}+r\}
-\max\{z_{\mathrm{lower},A_l},Z_{T_l}-r\}]$ denotes the length of the interval in which another PA would be closer to the target than the selected PA located at ${Z}_{\hat{P}_l}$. 

\vspace{-.85em}
\subsection{\textbf{Outage Analysis}}
For mathematical tractability, we consider a two-user NOMA case. The users associated with the $A_l$ segment are statistically ordered according to their average effective channel powers such that 
$\mathbb{E}\left[\left|h_{1,1}^{1,\mathrm{eff}}\right|^2\right]>\mathbb{E}\left[\left|h_{1,2}^{1,\mathrm{eff},}\right|^2\right]$. Let $\boldsymbol{\omega}$ denote a realization of the user location,
where $\boldsymbol{\omega}=[x,y,0]^{\mathrm T}$ for a CU and $\boldsymbol{\omega}=[x,y,z]^{\mathrm T}$ for an AU. The average effective channel power at $\boldsymbol{\omega}$ is as
${H}_{l}(\boldsymbol{\omega})
=\mathbb{E}\!\left[\left|h_{l}^{l,\mathrm{eff}}\right|^2\middle|\boldsymbol{\Omega}=\boldsymbol{\omega}\right].$
Let $F_{H_l}(\cdot)$ denote the CDF of the average effective channel
power $H_l(\boldsymbol{\Omega})$ of an unordered user, defined as $ F_{H_l}(x)
=\Pr\left\{H_l(\boldsymbol{\Omega})\le x\right\}$.
For $K=2$, the spatial PDFs of the statistically ordered strong and weak users are given, respectively, by
\begin{equation}
f_{\boldsymbol{\Omega}_{l,1}}(\boldsymbol{\omega})
=
2f_{\boldsymbol{\Omega}}(\boldsymbol{\omega})
F_{H_l}\!\left(H_l(\boldsymbol{\omega})\right),
\label{eq:strong users}
\end{equation}
\begin{equation}
f_{\boldsymbol{\Omega}_{l,2}}(\boldsymbol{\omega})
=2f_{\boldsymbol{\Omega}}(\boldsymbol{\omega})
\left[1-F_{H_l}\!\left(H_l(\boldsymbol{\omega})\right)
\right]. 
\label{eq:weak users}
\end{equation}

\subsubsection{\textbf{Ground Users}}
$CU_{1,1}$ and $CU_{1,2}$ denote the statistically strong and weak users, respectively. The corresponding thresholds are given as $\gamma_{1,1}^{\mathrm{th}}=2^{R_{1,1}^{\mathrm{th}}}-1$ and $\gamma_{1,2}^{\mathrm{th}}=2^{R_{1,2}^{\mathrm{th}}}-1$, where $R_{1,1}^{\mathrm{th}}$ and $R_{1,2}^{\mathrm{th}}$ are the threshold rates for $CU_{1,1}$ and $CU_{1,2}$, respectively.
\paragraph{\textbf{Outage Probability of the Weak User} }
The weak user directly decodes its own information signal while treating the signal intended for strong users as interference. By using \eqref{eq:eq18} the  instantaneous SINR of weak users is given as 
\vspace{-0.0em}
\begin{equation}
\gamma_{1,2}
=\frac{\mathcal{P}_{1_{m}}\left\lvert h_{1,2}^{1,\mathrm{eff}}\right\rvert^2
\alpha b_{1,2}}{I_{intra}+\mathcal{P}_{2_{m}}\left\lvert h_{1,2}^{2,\mathrm{eff}}\right\rvert^2+\sigma_{1,2}^2}.
\end{equation}
The outage probability of $CU_{1,2}$ is defined as 
$P_{1,2}^{\mathrm{out}} = \Pr \left( \gamma_{1,2} <\gamma_{1,2}^{\mathrm{th}} \right)$. The successful decoding condition for  the  signal of $CU_{1,2}$ is  
$\gamma_{1,2}\geq\gamma_{1,2}^{\mathrm{th}}$ 
can be rearranged as $
\mathcal{P}_{1_m} \Big[ \alpha b_{1,2} 
- \gamma_{1,2}^{\mathrm{th}} \left( \alpha b_{1,1}+1-\alpha \right) \Big] \left\lvert h_{1,2}^{1,\mathrm{eff}}\right\rvert^2
\geq \gamma_{1,2}^{\mathrm{th}} \left(\mathcal{P}_{1_m}\left\lvert h_{1,2}^{2,\mathrm{eff}}\right\rvert^2 + \sigma_{1,2}^{2} \right)$.  
Let $\Delta_{l,2}=\alpha b_{l,2} -\gamma_{l,2}^{\mathrm{th}}\left(\alpha b_{l,1}+1-\alpha\right)$. For successful decoding of $CU_{1,2}$, $\Delta_{1,2}>0$ and also define $\mathcal{T}_{1,2}=\frac{\gamma_{1,2}^{\mathrm{th}}}{\Delta_{1,2}}$. Thus, the weak-user outage becomes  
\begin{equation} 
P_{1,2}^{\mathrm{out}} 
= \Pr \left[ \left\lvert h_{1,2}^{1,\mathrm{eff}}\right\rvert^2\leq\mathcal{T}_{1,2} \left( 
\left\lvert h_{1,2}^{2,\mathrm{eff}}\right\rvert^2
+ \frac{\sigma_{1,2}^{2}}{\mathcal{P}_{1_m}} \right) \right]. 
\label{eq:weak_channel_outage} 
\end{equation}
By using \eqref{eq:effective_power_cdf}, rearranging and conditioning on interference channel  power \eqref{eq:weak_channel_outage} is given as  \vspace{-0.0em}
\begin{equation} 
P_{1,2}^{\mathrm{out}}\mid{H_{1,2}^{2,\mathrm{eff}}}
= 
F_{H_{1,2}^{1,\mathrm{eff}}}
(\left[ 
\mathcal{T}_{1,2} 
\left( 
{H_{1,2}^{2,\mathrm{eff}}}+ 
\frac{\sigma_{1,2}^{2}}{\mathcal{P}_{1_m}} 
\right)\mid\boldsymbol{\Omega}_{CU_{1,2}})
\right]. 
\label{eq:weak_conditional_outage} 
\end{equation} 
Using \eqref{eq:effective_power_cdf} and \eqref{eq:weak users} averaging over the $CU_{1,2}$ distribution, the outage probability of the weak user is given by  \vspace{-0.0em}
\begin{equation}
\begin{split}
P_{1,2}^{\mathrm{out}}
&=
\frac{2}{XY}
\int_{-\frac{X}{2}}^{\frac{X}{2}}
\int_{-\frac{Y}{2}}^{\frac{Y}{2}}
\int_{0}^{\infty}
F_{H_{1,2}^{1,\mathrm{eff}}}
\\&\hspace{0mm}
\left[
\mathcal{T}_{1,2}
\left(
\iota+
\frac{\sigma_{1,2}^{2}}{\mathcal{P}_{1_m}}
\right)\mid
\boldsymbol{\Omega}_{CU_{1,2}}
\right]
f_{H_{1,2}^{2,\mathrm{eff}}}
\left(
\iota\mid
\boldsymbol{\Omega}_{CU_{1,2}}\right)
\\&\hspace{0mm}
\left[
1-
F_{H_1}
\left(
H_1(\boldsymbol{\omega}_{CU_{1,2}})
\right)
\right]
\,d\iota\,dy_{CU_{1,2}}\,dx_{CU_{1,2}}.
\end{split}
\label{eq:weak_outage_random_users}
\end{equation}
where $\iota={H_{1,2}^{2,\mathrm{eff}}}$.
\vspace{-0.5em}
\paragraph{\textbf{Outage Probability of the Strong User}}
The strong user first decodes the higher-power signal intended for the weak user and subsequently performs SIC before decoding its own information signal. Therefore, successful reception at $CU_{1,1}$ requires both $\gamma_{1,1\rightarrow2}\geq\gamma_{1,2}^{\mathrm{th}}$ and $\gamma_{1,1}\geq\gamma_{1,1}^{\mathrm{th}}$. 
The SINR at $CU_{1,1}$ for decoding the signal intended for $CU_{1,2}$ is given by \vspace{-0.0em} 
\begin{equation} 
\gamma_{1,1\rightarrow2} = \frac{\mathcal{P}_{1_{m}}\left\lvert h_{1,1}^{1,\mathrm{eff}}\right\rvert^2
\alpha b_{1,2}}{\mathcal{P}_{1_{m}}\left\lvert h_{1,1}^{1,\mathrm{eff}}\right\rvert^2\left( 
\alpha b_{1,1}+1-\alpha \right)+\mathcal{P}_{2_{m}}\left\lvert h_{1,1}^{2,\mathrm{eff}}\right\rvert^2+\sigma_{1,1}^2}, 
\label{eq:strong_decode_weak_sinr} 
\end{equation} 
rearranging 
$\gamma_{1,1\rightarrow2}\geq\gamma_{1,2}^{\mathrm{th}}$ gives the corresponding threshold \vspace{-0.5em}
\begin{equation} 
\mathcal{T}_{1,1\rightarrow2} 
= \frac{ \gamma_{1,2}^{\mathrm{th}} }{ 
\alpha b_{1,2} - \gamma_{1,2}^{\mathrm{th}} \left( 
\alpha b_{1,1}+1-\alpha \right) }. 
\label{eq:strong_decode_weak_threshold} 
\end{equation} 
After decoding and successfully cancelling the signal intended for $CU_{1,2}$, the SINR for decoding the desired signal of $CU_{1,1}$ is given by 
\vspace{-0.5em}
\begin{equation} 
\gamma_{1,1} = \frac{\mathcal{P}_{1_{m}}\left\lvert h_{1,1}^{1,\mathrm{eff}}\right\rvert^2
\alpha b_{1,1}}{\mathcal{P}_{1_{m}}\left\lvert h_{1,1}^{1,\mathrm{eff}}\right\rvert^2
(1-\alpha)+\mathcal{P}_{2_{m}}\left\lvert h_{1,1}^{2,\mathrm{eff}}\right\rvert^2+\sigma_{1,1}^2}. 
\label{eq:strong_own_sinr} 
\end{equation} 
Similarly, rearranging $\gamma_{1,1}\geq\gamma_{1,1}^{\mathrm{th}}$ 
gives $\mathcal{T}_{1,1} =\frac{\gamma_{1,1}^{\mathrm{th}}}
{\alpha b_{1,1}-\gamma_{1,1}^{\mathrm{th}}(1-\alpha)}$. Since the same desired channel gain ${H_{1,1}^{1,\mathrm{eff}}}$ must satisfy both decoding requirements, the effective SIC decoding threshold is $
\mathcal{T}_{1,1}^{\mathrm{SIC}}=\max
\left\{\mathcal{T}_{1,1\rightarrow2},\mathcal{T}_{1,1}
\right\}$. Thus, the outage probability of the strong user can be written as 
\begin{equation} 
P_{1,1}^{\mathrm{out}} 
= \Pr \left[ {H_{1,1}^{1,\mathrm{eff}}}\leq \mathcal{T}_{1,1}^{\mathrm{SIC}} \left({H_{1,1}^{2,\mathrm{eff}}} 
+ \frac{\sigma_{1,1}^{2}}{\mathcal{P}_{1_m}} \right) \right], 
\label{eq:strong_channel_outage} 
\end{equation} 
Using \eqref{eq:effective_power_cdf} and \eqref{eq:strong users}, averaging over the $CU_{1,1}$distribution, the outage probability of the strong user is given by  
\begin{equation}
\begin{split}
P_{1,1}^{\mathrm{out}}
&=
\frac{2}{XY}
\int_{-\frac{X}{2}}^{\frac{X}{2}}
\int_{-\frac{Y}{2}}^{\frac{Y}{2}}
\int_{0}^{\infty}
F_{H_{1,1}^{1,\mathrm{eff}}}
\\&\hspace{0mm}
\left[
\mathcal{T}_{1,1}^{\mathrm{SIC}}
\left(
\iota+
\frac{\sigma_{1,1}^{2}}{\mathcal{P}_{1_m}}
\right)
\middle|
\boldsymbol{\Omega}_{CU_{1,1}}
\right]
f_{H_{1,1}^{2,\mathrm{eff}}}
\left(
\iota
\middle|
\boldsymbol{\Omega}_{CU_{1,1}}\right)
\\&\hspace{0mm}
F_{H_1}
\left(
H_1(\boldsymbol{\omega}_{CU_{1,1}})
\right)
\,d\iota\,dy_{CU_{1,1}}\,dx_{CU_{1,1}},
\end{split}
\label{eq:strong_outage_random_users}
\end{equation}
where $\iota={H_{1,1}^{\mathrm{eff},2}}$ .
\subsubsection{{\textbf{Outage Probability of Aerial User}}}
$AU_{2,1}$ and $AU_{2,2}$ denote the statistically strong and weak users, respectively. The corresponding thresholds are given as $\gamma_{2,1}^{\mathrm{th}}=2^{R_{2,1}^{\mathrm{th}}}-1$ and $\gamma_{2,2}^{\mathrm{th}}=2^{R_{2,2}^{\mathrm{th}}}-1$, where $R_{2,1}^{\mathrm{th}}$ and $R_{2,2}^{\mathrm{th}}$ are the threshold rates for $AU_{2,1}$ and $AU_{2,2}$, respectively.

\paragraph{ \textbf{Outage Probability of the Weak AU}}
 Similar to the ground case, the weak AU directly decodes its own information signal while treating the signal intended for the strong AU as intra-segment interference. Using \eqref{eq:eq19} instantaneous SINR is given as \vspace{-0.5em}
\begin{equation}
\gamma_{2,2}
=\frac{
\mathcal{P}_{2_m}\left\lvert h_{2,2}^{2,\mathrm{eff}}\right\rvert^2
\alpha b_{2,2}}{I_{intra}+\mathcal{P}_{1_{m}}\left\lvert h_{2,2}^{1,\mathrm{eff}}\right\rvert^2+\mathcal{P}_{3_{m}}\left\lvert h_{2,2}^{3,\mathrm{eff}}\right\rvert^2+\sigma_{2,2}^2}.
\label{eq:au_weak_sinr}
\end{equation}
The outage probability is defined as $
P_{2,2}^\mathrm{out}=\Pr\left(\gamma_{2,2}<\gamma_{2,2}^{\mathrm{th}}\right)$.
Rearranging the successful decoding condition gives
\vspace{-0.25em}
\begin{equation}
H_{2,2}^{2,\mathrm{eff}}\geq\mathcal{T}_{2,2}\left(H_{2,2}^{1,\mathrm{eff}}+H_{2,2}^{3,\mathrm{eff}}+\frac{\sigma_{2,2}^{2}}{\mathcal{P}_{2_m}}\right),
\label{eq:au_weak_channel_condition}
\end{equation}
where $\mathcal{T}_{2,2}=\frac{\gamma_{2,2}^{\mathrm{th}}}{\alpha b_{2,2}-\gamma_{2,2}^{\mathrm{th}}
\left(\alpha b_{2,1}+1-\alpha\right)}$,
provided that $ \alpha b_{2,2}-\gamma_{2,2}^{\mathrm{th}}\left(\alpha b_{2,1}+1-\alpha\right)>0$. Therefore, 
\vspace{-0.5em}
\begin{equation}
P_{2,2}^{\mathrm{out}}=\Pr\left[H_{2,2}^{2,\mathrm{eff}}\leq\mathcal{T}_{2,2}
\left(H_{2,2}^{1,\mathrm{eff}}+H_{2,2}^{3,\mathrm{eff}}+\frac{\sigma_{2,2}^{2}}{\mathcal{P}_{2_m}}\right)
\right].
\label{eq:au_weak_channel_outage}
\end{equation}
Since the interference channel-power gains from the two adjacent segments are conditionally independent for a given AU position, by using \eqref{eq:au_pdf} their joint PDF can be expressed as
$f_{H_{2,2}^{1,\mathrm{eff}},H_{2,2}^{3,\mathrm{eff}}}
\left(\iota_1,\iota_2\mid\boldsymbol{\Omega}_{AU_{2,2}}
\right)=f_{H_{2,2}^{1,\mathrm{eff}}}
\left(\iota_1\mid\boldsymbol{\Omega}_{AU_{2,2}}\right)f_{H_{2,2}^{3,\mathrm{eff}}}\left(\iota_2\mid\boldsymbol{\Omega}_{AU_{2,2}}\right)$.
conditioned on the interference realisations
${H_{2,2}^{1,\mathrm{eff}}}=\iota_1$ and ${H_{2,2}^{3,\mathrm{eff}}}=\iota_2$, and using \eqref{eq:au_cdf} the conditional  outage probability of the weak AU is \vspace{-0.5em}
\begin{equation}
P_{2,2}^{\mathrm{out}}
\big|\iota_1,\iota_2\!=\!F_{H_{2,2}^{2,\mathrm{eff}}}\left[
\mathcal{T}_{2,2}\left(\iota_1+\iota_2+
\frac{\sigma_{2,2}^{2}}{\mathcal{P}_{2_m}}\right)\mid\boldsymbol{\Omega}_{AU_{2,2}}\right]\!.
\label{eq:au_weak_conditional_outage}
\end{equation}
Using \eqref{eq:au_cdf} and \eqref{eq:weak users} averaging over the $AU_{2,2}$ distribution, the outage probability of the weak user is given by \eqref{eq:au_weak_outage_final}.
\begin{figure*}[!t]
\begin{equation}
\begin{split}
P_{2,2}^{\mathrm{out}}
&=
\frac{2}{XYA_{2,\mathrm{seg}}}
\int_{-\frac{X}{2}}^{\frac{X}{2}}
\int_{-\frac{Y}{2}}^{\frac{Y}{2}}
\int_{A_{2,\mathrm{seg},\min}}^{A_{2,\mathrm{seg},\max}}
\int_{0}^{\infty}
\int_{0}^{\infty}
F_{H_{2,2}^{2,\mathrm{eff}}}
\left[
\mathcal{T}_{2,2}
\left(
\iota_1+\iota_2+
\frac{\sigma_{2,2}^{2}}{\mathrm{P}_{2_m}}
\right)\mid\boldsymbol{\Omega}_{AU_{2,2}}
\right]\times
\\&\hspace{30mm}
f_{H_{2,2}^{1,\mathrm{eff}}}(\iota_1\mid
\boldsymbol{\Omega}_{AU_{2,2}})
f_{H_{2,2}^{3,\mathrm{eff}}}(\iota_2\mid
\boldsymbol{\Omega}_{AU_{2,2}})
\left[
1-
F_{H_2}
\left(
H_2(\boldsymbol{\omega}_{AU_{2,2}})
\right)
\right]
d\iota_1\,d\iota_2\,
dz_{AU_{2,2}}\,
dy_{AU_{2,2}}\,
dx_{AU_{2,2}} .
\end{split}
\label{eq:au_weak_outage_final}
\end{equation}
\vspace{-1.25em}
\end{figure*}
\paragraph{\textbf{Outage Probability of the Strong AU}} 
The strong aerial user first decodes the higher power signal intended for weak users and subsequently performs SIC before decoding its own information signal. Therefore, successful decoding of own signal requires  $\gamma_{2,1\rightarrow2}\geq\gamma_{l,2}^{\mathrm{th}},\gamma_{2,1}\geq\gamma_{2,1}^{\mathrm{th}}$. The SINR at $AU_{2,1}$ for decoding the signal intended for $AU_{2,2}$ is
\begin{equation}
\gamma_{2,1\rightarrow2}=
\frac{
\mathcal{P}_{2_m}\left\lvert h_{2,1}^{2,\mathrm{eff}}\right\rvert^2
\alpha b_{2,2}}{\upsilon+\mathcal{P}_{1_{m}}\left\lvert h_{2,1}^{1,\mathrm{eff}}\right\rvert^2+\mathcal{P}_{3_{m}}\left\lvert h_{2,1}^{3,\mathrm{eff}}\right\rvert^2+\sigma_{2,1}^2},
\label{eq:au_strong_decode_weak_sinr}
\end{equation}
 where $\upsilon=\mathcal{P}_{2_{m}}\left\lvert h_{2,1}^{2,\mathrm{eff}}\right\rvert^2\left( 
\alpha b_{2,1}+1-\alpha \right)$. By rearranging $\gamma_{2,1\rightarrow2}\geq\gamma_{2,2}^{\mathrm{th}}$, gives
$\mathcal{T}_{2,1\rightarrow2}=\frac{\gamma_{2,2}^{\mathrm{th}}
}{\alpha b_{2,2}-\gamma_{2,2}^{\mathrm{th}}\left(\alpha b_{2,1}+1-\alpha\right)}$.
After decoding the weak signal and  successfully cancelling it by SIC, the SINR for decoding the strong user signal  becomes
\vspace{-0.65em}
\begin{equation}
\gamma_{2,1}=\frac{
\mathcal{P}_{2_m}\left\lvert h_{2,1}^{2,\mathrm{eff}}\right\rvert^2
\alpha b_{2,1}}{\mathcal{P}_{2_{m}}\left(\left\lvert h_{2,1}^{2,\mathrm{eff}}\right\rvert^2\left(1-\alpha\right)+\left\lvert h_{2,1}^{1,\mathrm{eff}}\right\rvert^2+\left\lvert h_{2,1}^{3,\mathrm{eff}}\right\rvert^2\right)+\sigma_{2,1}^2}.
\label{eq:au_strong_own_sinr}
\end{equation}
The corresponding channel gain threshold is
$ \mathcal{T}_{2,1}=\frac{\gamma_{2,1}^{\mathrm{th}}
}{\alpha b_{2,1}-\gamma_{2,1}^{\mathrm{th}}(1-\alpha)
}$. Since the same effective channel gain $H_{2,1}^{2,\mathrm{eff}}$ must satisfy both decoding requirements, the effective SIC threshold is $ \mathcal{T}_{2,1}^{\mathrm{SIC}}
=\max\left\{\mathcal{T}_{2,1\rightarrow2},
\mathcal{T}_{2,1}\right\}$. Thus, the strong-AU outage probability is given as \vspace{-0.5em}
\begin{equation}
P_{2,1}^{\mathrm{out}}=\Pr
\left[H_{2,1}^{2,\mathrm{eff}}\leq\mathcal{T}_{2,1}^{\mathrm{SIC}}
\left(H_{2,1}^{1,\mathrm{eff}}+H_{2,1}^{3,\mathrm{eff}}+\frac{\sigma_{2,1}^{2}}{\mathcal{P}_{2_m}}\right)\right].
\label{eq:au_strong_channel_outage}
\end{equation}
Consequently, by using \eqref{eq:au_pdf} and \eqref{eq:au_cdf} the outage probability of the strong AU is given as 
\vspace{-0.5em}
\begin{equation}
P_{2,1}^{\mathrm{out}}\mid\iota_1,\iota_2
= F_{H_{2,1}^{2,\mathrm{eff}}}
\left[\mathcal{T}_{2,1}^{\mathrm{SIC}}\left(
\iota_1+\iota_2+\frac{\sigma_{2,1}^{2}}{\mathcal{P}_{2_m}}\right)\mid\boldsymbol{\Omega}_{AU_{2,1}}\right] .
\label{eq:au_strong_outage_integral}
\end{equation}
Using \eqref{eq:au_cdf} and \eqref{eq:strong users} averaging over the $AU_{2,1}$ distribution, the outage probability of the strong user is given by  is given by \eqref{eq:au_strong_outage_final}.
\begin{figure*}[!t]
\begin{equation}
\begin{split}
P_{2,1}^{\mathrm{out}}
&=
\frac{2}{XYA_{2,\mathrm{seg}}}
\int_{-\frac{X}{2}}^{\frac{X}{2}}
\int_{-\frac{Y}{2}}^{\frac{Y}{2}}
\int_{A_{2,\mathrm{seg},\min}}^{A_{2,\mathrm{seg},\max}}
\int_{0}^{\infty}
\int_{0}^{\infty}
F_{H_{2,1}^{2,\mathrm{eff}}}
\left[
\mathcal{T}_{2,1}^{\mathrm{SIC}}
\left(
\iota_1+\iota_2+
\frac{\sigma_{2,1}^{2}}{\mathcal{P}_{2_m}}
\right)\mid\boldsymbol{\Omega}_{AU_{2,1}}
\right]\times
\\&\hspace{40mm}
f_{H_{2,1}^{1,\mathrm{eff}}}(\iota_1\mid
\boldsymbol{\Omega}_{AU_{2,1}})
f_{H_{2,1}^{3,\mathrm{eff}}}(\iota_2\mid
\boldsymbol{\Omega}_{AU_{2,1}})
F_{H_2}
\left(
H_2(\boldsymbol{\omega}_{AU_{2,1}})
\right)
d\iota_1\,d\iota_2\,
dz_{AU_{2,1}}\,
dy_{AU_{2,1}}\,
dx_{AU_{2,1}}.
\end{split}
\label{eq:au_strong_outage_final}
\end{equation}
\vspace{-1.25em}
\end{figure*}
\vspace{-0.5em}

The above outage expressions are valid for positive decoding-threshold denominators. Otherwise, the corresponding user outage probability is $1$.
\subsubsection{\textbf{ Segment Outage Probability}}
The segment is in outage if either of the two users is in outage. Thus, the segment outage probability is given by
\begin{equation}
P_{l,seg}^{\mathrm{out}}
=P_{l,1}^{\mathrm{out}}
+P_{l,2}^{\mathrm{out}}
-\Pr\left(\mathcal{O}_{l,1}\cap\mathcal{O}_{l,2}
\right),
\label{eq:segment_outage_general}
\end{equation}
where $\mathcal{O}_{l,1}$ and $\mathcal{O}_{l,2}$ denote the outage events of the strong and weak users, respectively. Assuming independent user outage events \cite{S3}, the segment outage probability is obtained from the product of the individual user success probabilities and \eqref{eq:segment_outage_general} reduces to $P_{l,\mathrm{seg}}^{\mathrm{out}}
=1-\left(1-P_{l,1}^{\mathrm{out}}\right)\left(1-P_{l,2}^{\mathrm{out}}\right)$.
\subsubsection{{\textbf{Sensing Outage Probability}}}
The communication outage probability is not directly applicable to sensing, since sensing performance is governed by estimation accuracy rather than the achievable rate. For a unified evaluation of communication and sensing performance, the
sensing outage probability is defined in terms of the MSE, following \cite{S3,S5}. In this work, the sensing MSE represents the target-range estimation error, where the target range is obtained from the round-trip delay $\tau$ of the received reflected sensing signal. Accordingly, a sensing outage occurs when the MSE exceeds a given threshold, $P_l^{\mathrm{out,sen}}=\Pr\left(\rho_s>\rho_{\mathrm{Th}}\right)$, where $\rho_s$ denotes the sensing MSE and $\rho_{\mathrm{Th}}$ is the target MSE threshold. For a sensing signal with a flat spectral density, the CRLB for range estimation is given by $\rho_s \geq \rho_s^{\mathrm{CRLB}}=\frac{c^2}{K_S\gamma_l^{\mathrm{sen}}\left(\beta B\right)^2}$, where $K_S=\frac{64\pi^2}{12}$, $\beta$ is the spectrum allocation factor for sensing, and $B$ is the bandwidth. Assuming an efficient estimator that achieves the CRLB, the sensing outage probability is expressed in terms of the sensing SINR as $P_l^{\mathrm{out,sen}}=\Pr\left(\gamma_l^{\mathrm{sen}}\leq\tau_s\right)$, where $\tau_s=\frac{c^2}{K_S\rho_{\mathrm{Th}}\left(\beta B\right)^2}$ denotes the required sensing SINR threshold. Since the interference forward  channel-power gains from the two adjacent segments are conditionally independent for a given ST position, their joint PDF can be expressed as
$f_{H_{2,T}^{\mathrm{eff},1},H_{2,T}^{\mathrm{eff},3}}
(\iota_1,\iota_2)=f_{{H_{2,T}^{\mathrm{eff},1}}}(\iota_1)f_{H_{2,T}^{\mathrm{eff},3}}(\iota_2)$. Conditioned on the interference realisations
${H_{2,T}^{\mathrm{eff},1}}=\iota_1$ and
${H_{2,T}^{\mathrm{eff},3}}=\iota_2$, the common backward channel gain $\widetilde{H}_{l,\mathrm{sen}}$ is deterministic, since the receiving PA is selected and fixed for a given ST position, as described in \eqref{eq:nearest_pa_pdf} . Using the SINR value from \eqref{eq:eq22} and the PDF and CDF of the forward channel from \eqref{eq:sen_forward_pdf}, \eqref{eq:sen_forward_cdf}, the sensing outage is  given by \eqref{eq:sensing_outage}  \vspace{-0.5em}
\begin{figure*}[!t]
 \begin{equation}
\begin{aligned}
P_{2}^{\mathrm{out,sen}}=&\frac{1}{XYA_{2,\mathrm{seg}}}\int_{-X/2}^{X/2}\int_{-Y/2}^{Y/2}\int_{A_{2,\mathrm{seg},\min}}^{A_{2,\mathrm{seg},\max}}\int_{A_{2,\mathrm{seg},\min}}^{A_{2,\mathrm{seg},\max}}\int_{0}^{\infty}\int_{0}^{\infty}
\\&\hspace{20mm}F_{H_{2,{T_2}}^{2,\mathrm{eff}}}\left(\frac{\tau_s}{\mathcal{P}_{2_m}(1-\alpha)}\left[\mathcal{P}_{2_m}\alpha(\iota_1+\iota_2)+\frac{\sigma_{2,s}^{2}}{\widetilde{H}_{2,\mathrm{sen}}\left(Z_{\hat{P}_{2,m}},\boldsymbol{\Omega}_{T_2}\right)}\right]\mid\boldsymbol{\Omega}_{T_2}\right)\times\\&\hspace{35mm}f_{H_{2,{T_2}}^{1,\mathrm{eff}}}\left(\iota_1\mid\boldsymbol{\Omega}_{T_2}\right)f_{H_{2,T_2}^{3,\mathrm{eff}}}\left(\iota_2\mid\boldsymbol{\Omega}_{T_2}\right)f_{Z_{\hat{P}_{2,m}}}\left(Z_{\hat{P}_{2,m}}\mid Z{T_2}\right)\,d\iota_1\,d\iota_2\,dz_{\hat{P}_{2,m}}\,dz_{T_2}\,dy_{T_2}\,dx_{T_2} .
\end{aligned}
\label{eq:sensing_outage}
\end{equation}
\hrule
\end{figure*}.
\vspace{-0.5em}
\section{Results and Discussion}
\label{sec:result_discussion}
\vspace{-0.25em}
This section presents the performance of the proposed V-SWAN-assisted downlink NOMA-ISAC EDMA framework, which is evaluated to demonstrate its operational behaviour in different environmental conditions. The simulation results are presented and compared with analytical results to validate the accuracy of the proposed method. The analytical results closely match the simulation results for all the presented outage figures. In addition to this, the results of the proposed framework are also compared with existing frameworks like H-SWAN, V-PASS, and other state-of-the-art methods like OMA. For this, different simulation parameters are used, which are listed in Table \ref{Parameters}. Throughout the results discussion, the analytical results are denoted by \say{Ana}. The proposed framework jointly performs sensing and communication. Therefore, the performance evaluation of the communication and sensing functionalities is first discussed separately, followed by an evaluation of their joint ISAC performance. Accordingly, this section is divided into three sub-sections.

\begin{table}[t]
 \caption{ Simulation Parameter}
\label{Parameters}
\renewcommand{\arraystretch}{1.2}
\centering
 {\rowcolors{3}{white!100!white!100}{white!100!white!40}
\begin{tabular}{|c|c|}
\hline
 \rowcolor{lightgray!80}\textbf{Parameter} & \textbf{Value } \\
\hline
\rowcolor{Lavender!20}\multicolumn{2}{|c|}{\bf General System Parameter} \\
\hline
Transmit power ($\mathcal{P}_t$) &  $0-30$ dBm  \\
Carrier frequency ($f_c$) & $10$ GHz \\
Total serving area & $10$ m$\times10$ m$\times30$ m  \\
ISAC power splitting factor ($\alpha,1-\alpha$) & $0.7,0.3$\\
Background noise ($\sigma_{l,k}^{2}=\sigma_{l,s}^{2}$) & $10^{-10}$ W\\
Number of segments ($L$) & $3$\\
Number of PAs in each segment ($M$) & $8$\\
Effective refractive index ($\eta$) & $1.4$\\
In-waveguide attenuation coefficient & $0.05$ dB/m\\
Number of Monte Carlo simulations & $10^{5}$ \\
\hline
\rowcolor{Apricot!20}\multicolumn{2}{|c|}{\bf Communication  Parameter} \\
\hline
 Number of CU in $A_1$ and AU in $A_2$, respectively & $2,2$\\
 Path loss exponent $\nu$ (ground, aerial) & $2.5,2$ \\
 Target data rate $\left({R_{l,k}^{\mathrm{th}}}\right)$ & $0.5$ bps/Hz\\
 NOMA power allocation factor ($b_{l,k} $) & $0.4,0.6$\\
\hline
\rowcolor{Cyan!20}\multicolumn{2}{|c|}{\bf Sensing Parameter} \\
\hline
Radar Cross-Section ($\sigma_T$) & $2-20$ m${^2}$\\
CRLB parameter ($K_S$) & $\frac{64\pi^{2}}{12}$\\
Target MSE threshold ($\rho_{\mathrm{Th}}$) & 0.5 m$^2$  \\
Bandwidth ($B$) &  $30$ MHz\\ 
Spectrum allocation factor ($\beta$) & $\left[0,1\right]$\\ 
\hline
\end{tabular}}
\end{table}

\subsection{Communication Functionality}
\vspace{-0.25em}
In this subsection, we evaluate the communication performance of the proposed V-SWAN-assisted downlink NOMA-ISAC EDMA framework for two types of users, namely CUs and AUs, and compare it with other existing methods and frameworks like H-SWAN and V-PASS. For the CU, the proposed framework is denoted by V-SWAN NOMA-G, while the corresponding benchmark schemes are denoted by V-SWAN OMA-G, H-SWAN NOMA-G, H-SWAN OMA-G, V-PASS NOMA-G, and V-PASS OMA-G. Similarly, for the AU, the proposed framework is denoted by V-SWAN NOMA-A, while the corresponding  benchmark schemes are denoted by V-SWAN OMA-A, H-SWAN NOMA-A, and H-SWAN OMA-A.
\par
\begin{figure}[t]
    \centering
    \begin{subfigure}{0.49\linewidth}
        \centering
        \includegraphics[scale=0.38]{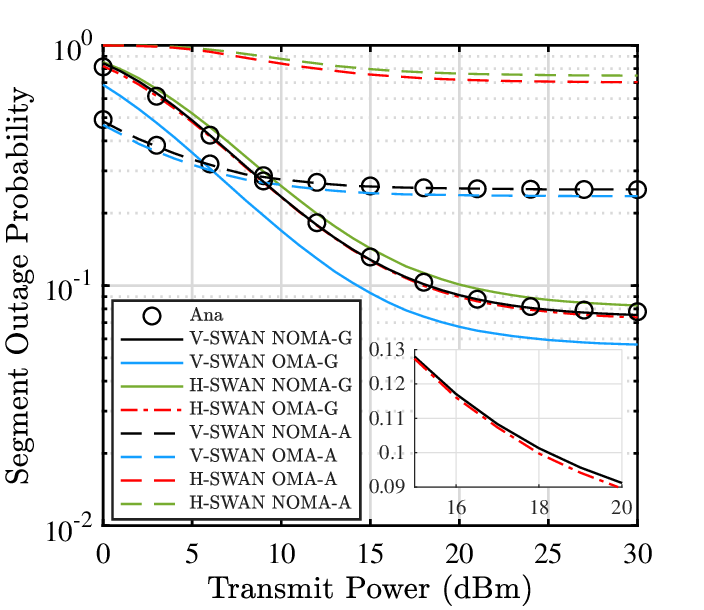}
        \caption{}
    \end{subfigure}
    \hfill
    \begin{subfigure}{0.49\linewidth}
        \centering
        \includegraphics[scale=0.38]{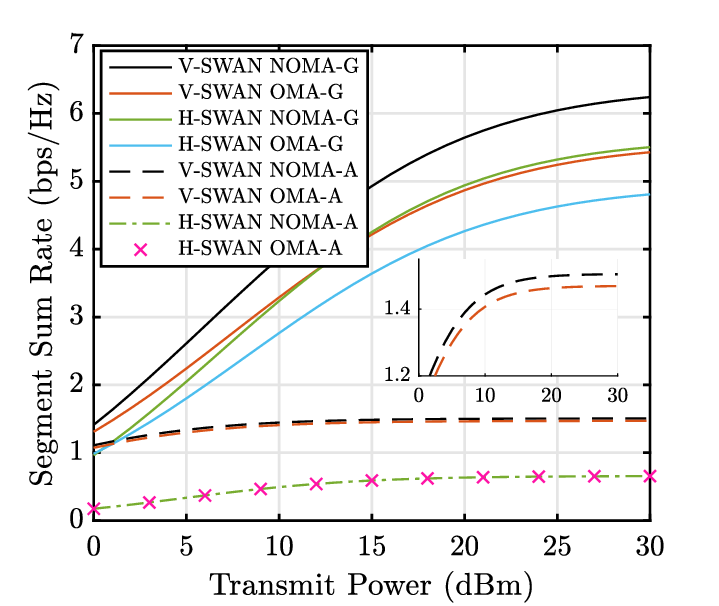}
    \caption{}
    \end{subfigure}
    \caption{(a) Segment outage v/s transmit power, and (b)  Segment sum rate  v/s transmit power.}
    \label{segment_power}
\end{figure}

In \figurename{~\ref{segment_power}}(a) the segment outage probability versus transmit power $\mathcal{P}_t$ is plotted. It can be observed that the segment outage of both users, CUs and AUs, decreases as $\mathcal{P}_t$ increases. But in the case of AU, inter-segment interference is more severe because there are two adjacent segments in comparison with CU. So, as the transmit power increases, the effect of thermal noise becomes progressively negligible, the outage probability continues to decrease initially, but at high transmit powers, both the desired signal power and the inter-segment interference power increase with a similar ratio, causing the system to gradually become interference-limited. So it starts to saturate at higher power, which can be seen in \figurename{~\ref{segment_power}}(a). Also, in the proposed V-SWAN, the PA positions vary along the vertical direction over the segment length. Due to this variation, the PA-to-CU distance decreases and consequently increases the effective channel gain, but in the case of H-SWAN, the waveguide is fixed at a distance, so the aerial distance between users and the PA is fixed, and due to this, the path loss is higher, so the segment outage of AUs is higher compared to CUs. The outage behaviour is further influenced by the multiple-access scheme. For the statistically strong user, OMA can achieve a lower outage probability than NOMA, since the OMA user is served without requiring SIC, whereas the strong NOMA user is assigned a smaller power coefficient, and its successful reception also depends on SIC. In contrast, the statistically weak NOMA user is allocated a larger power coefficient to compensate for its weaker channel condition. However, the resulting improvement in the weak user outage performance is not sufficient to compensate for the higher outage of the strong NOMA user. So, the overall segment outage probability of NOMA is higher than OMA.
\par
In \figurename{~\ref{segment_power}}(b), the segment rate for the proposed V-SWAN versus $\mathcal{P}_t$ is plotted and compared with different conventional PASS frameworks. It is observed that the segment rate increases for all the frameworks as $\mathcal{P}_t$ increases because SINR improves at higher transmit power of the decoding streams, thereby increasing the achievable rate of individual users, and thus the segment rate also increases. However, at higher power, the segment rate tends to saturate because the interference signal power also increases at a similar ratio to the desired signal power, making the system interference-limited. The AU segment rate is lower than the CU because, in the case of AU, there are two adjacent segments as compared to one in the CU case, and the segment rate saturates at a lower power than the CU. In OMA, the segment rate is lower than NOMA because NOMA uses power-domain multiplexing and simultaneously shares the same time and frequency resources. Therefore, the overall segment rate is higher for NOMA than OMA. This can also be observed in the figure, segment rates of NOMA and OMA for the AU are similar because interference is dominant. Therefore, providing a larger power coefficient to the weak user does not significantly affect the overall segment rate.
\par
\begin{figure}[t]
    \centering
    \begin{subfigure}{0.49\linewidth}
        \centering
        \includegraphics[scale=0.38]{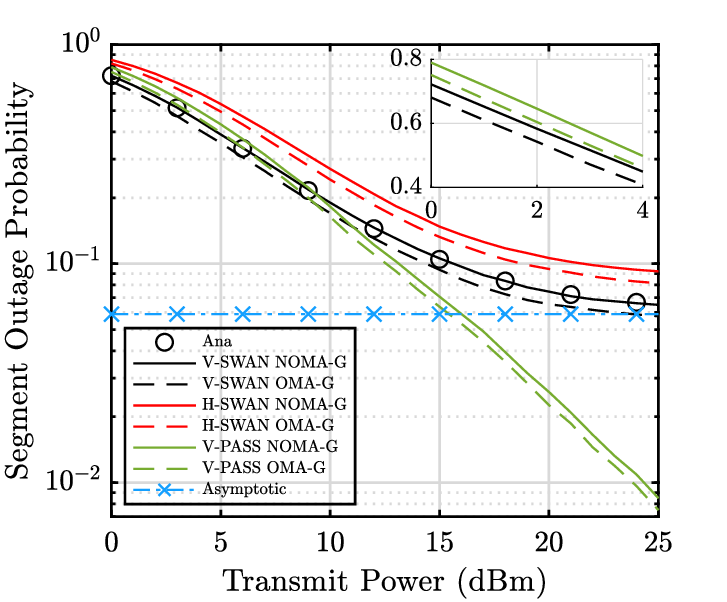}
        \caption{}
    \end{subfigure}
    \hfill
    \begin{subfigure}{0.49\linewidth}
        \centering
        \includegraphics[scale=0.38]{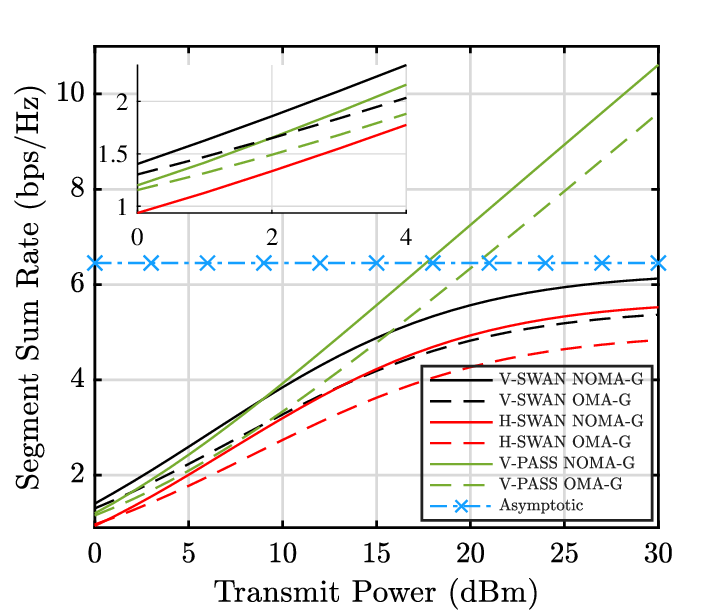}
    \caption{}
    \end{subfigure}
    \caption{(a) Segment outage v/s transmit power for different frameworks, and (b)  Segment sum rate  v/s transmit power for different frameworks.}
    \label{segment_framework}
\end{figure}

 In \figurename{~\ref{segment_framework}}(a), we further analyse the segment outage for different PASS frameworks and compare their performance with the proposed V-SWAN. Initially, the proposed V-SWAN performs well due to segmentation, which reduces the in-waveguide loss and also the distance between the users and PAs. However, at higher transmit power, the performance of V-SWAN degrades due to adjacent segment interference. On the other hand, V-PASS performs well at higher transmit power because it does not have adjacent segments. Therefore, it is not affected by adjacent segment interference. So, there is a trade-off between the proposed V-SWAN and conventional V-PASS. At lower 
transmit power, V-SWAN performs better, but when the transmit power exceeds $10$ dBm, V-PASS provides better performance.
\par
In \figurename{~\ref{segment_framework}}(b) segment rate versus  $\mathcal{P}_t$ is plotted for different conventional frameworks and compared with the proposed V-SWAN. This also follows the same trend. Initially, the V-SWAN segment rate is higher as compared to V-PASS due to segmentation, but later the V-PASS  segment rate is higher compared to the proposed one. One important observation in this plot is that the V-PASS segment  rate is not saturated like V-SWAN because there is no interfering segment in the case of V-PASS, so when $\mathcal{P}_t$ increases, the instantaneous rate of users increases, and so the segment rate also increases.  
\par Furthermore, for completeness, the asymptotic results are presented alongside the simulation results in \figurename{~\ref{segment_framework}}(a) and \figurename{~\ref{segment_framework}}(b), to provide further insight into the limiting behaviour of the proposed V-SWAN assisted downlink NOMA ISAC EDMA framework in the high transmit SNR $\left(\frac{\mathcal{P}_t}{\sigma_{l,k}}\to \infty\right)$ region. It is observed that, as the transmit power increases, both the segment outage probability and segment sum rate gradually converge to their respective asymptotic values, thereby highlighting the interference-limited nature of the considered framework.
\par
\begin{figure}[t]
    \centering
    \includegraphics[scale=0.45]{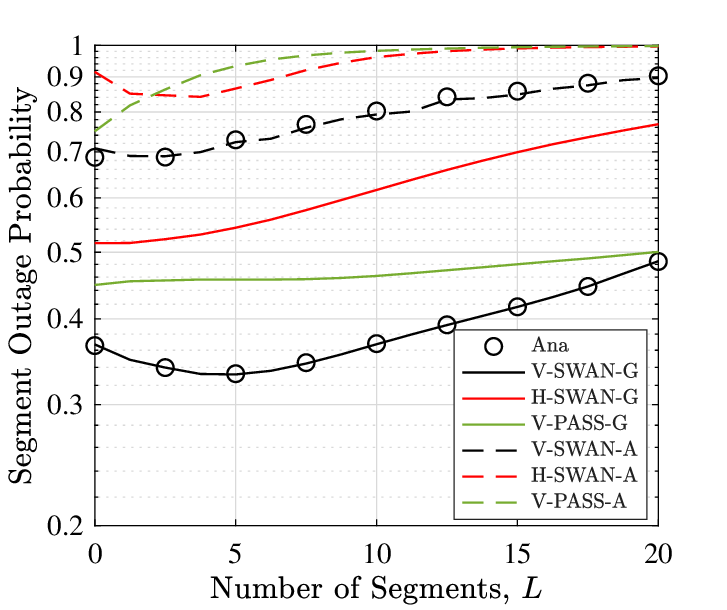}
    \caption{Segment outage with varying numbers of segments.}
    \label{segment-outge-Final}
\end{figure}

 \figurename{~\ref{segment-outge-Final}} illustrates the segment outage versus the number of segments $L$. Due to segmentation, the waveguide length is reduced, which reduces the distance between the FP and PA. As a result, the in-waveguide loss also reduces. Therefore, the segmented waveguide experiences lower in-waveguide loss compared with a single waveguide. As the number of segments increases, moving from a conventional single waveguide to SWAN, the segment outage initially decreases. However, after $L=5$, the segment outage starts increasing because the power radiated by each PA $\mathcal{P}_{l_m}$, decreases as the total transmit power is equally divided among the segments and then among the PAs within each segment. Hence, a U-shaped outage behaviour is observed. To clearly show this effect, we consider a fixed transmit power of $5$ dBm while varying the number of segments, with all other parameters remains unchanged. As expected, the segment outage initially decreases due to the reduced in-waveguide loss. However, for $L>5$, the reduction in the power available at each PA becomes dominant, increasing the segment outage. In the case of V-PASS, the outage increases with increasing $L$ because it uses a single waveguide with a larger length. As $L$ increases, V-PASS does not get the benefit of reduced in-waveguide loss as in SWAN, while the transmit power available at each PA decreases. Hence, the effect of lower transmit power becomes dominant, which increases the segment outage probability. 

\subsection{Sensing Functionality}
In this subsection, we discuss the sensing functionality of the proposed framework. For sensing, we consider a single target in each segment. For sensing, an additional backward path is considered, through which the reflected signal from the target is received at the selected PA and then propagated to the FP for sensing. The simulation results for sensing outage are obtained using \eqref{eq:eq22}, while the analytical sensing outage is obtained with the help of \eqref{eq:sensing_outage}.  The simulation results closely match the analytical results, validating the derived analytical expressions.
\par
\begin{figure}[t]
    \centering
    \begin{subfigure}{0.49\linewidth}
        \centering
        \includegraphics[scale=0.38]{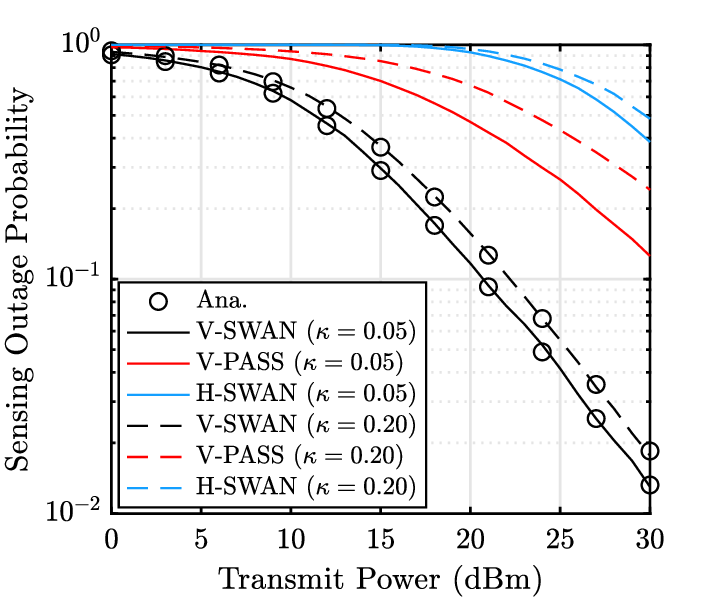}
        \caption{}
    \end{subfigure}
    \hfill
    \begin{subfigure}{0.49\linewidth}
        \centering
        \includegraphics[scale=0.38]{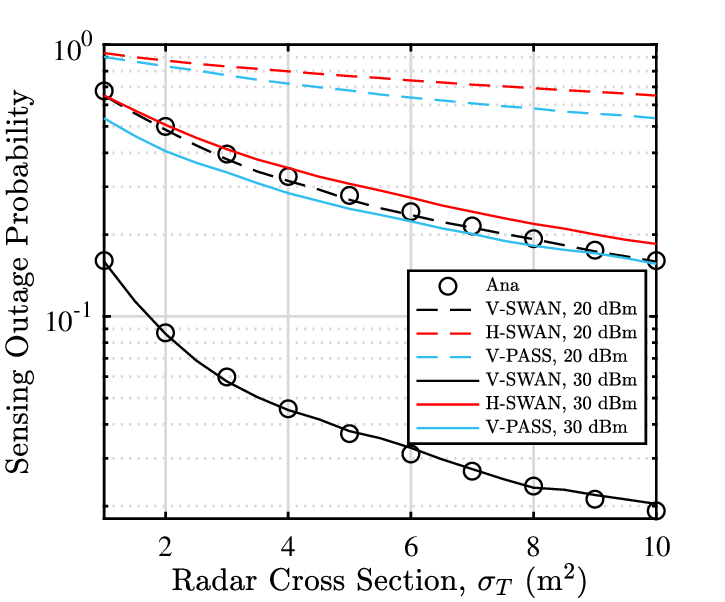}
    \caption{}
    \end{subfigure}
    \caption{(a) Sensing outage v/s transmit power, and (b) Sensing outage v/s transmit power with varying radar cross-section area.}
    \label{SensingOutage}
\end{figure}

 In \figurename{~\ref{SensingOutage}}(a), the sensing outage decreases as the transmit power $\mathcal{P}_t$ increases due to the improvement in the sensing SINR. The sensing outage is also strongly affected by the in-waveguide attenuation coefficient $\kappa$, since the in-waveguide attenuation affects both the forward and backward propagation paths of the sensing signal. Therefore, a larger value of $\kappa$ results in higher attenuation of the reflected sensing signal and, due to this, higher sensing outage. Moreover, the sensing signals from the adjacent segments are orthogonal to the serving segment sensing signal, while the communication interference from the same segment is removed at the sensing receiver. As a result, the interference in the sensing signal reception is significantly reduced, and the sensing SINR is less interference-limited as compared to the communication SINR as the transmit power increases. Hence, the sensing outage continues to decrease with increasing $\mathcal{P}_t$. It can also be observed that, for a smaller value of $\kappa$, the proposed framework achieves a lower sensing outage compared with the conventional frameworks.
\par In \figurename{~\ref{SensingOutage}}(b), the sensing outage is illustrated for different values of the radar cross section $\sigma_T$. The sensing outage decreases as $\sigma_T$ increases because a larger $\sigma_T$ strengthens the reflected signal and improves the sensing SINR. The sensing outage also strongly depends on the transmit power. So, at higher transmit power, more power is available for sensing, and the sensing outage decreases faster with increasing $\sigma_T$. Hence, the effect of the desired sensing signal becomes more dominant compared with the adjacent-segment interference, and the sensing outage decreases faster. Therefore, for  $30$ dBm transmit power, the proposed framework achieves a lower sensing outage compared to the other conventional frameworks.
\subsection{ISAC Performance}
\begin{figure}[t]
    \centering
    \includegraphics[scale=0.45]{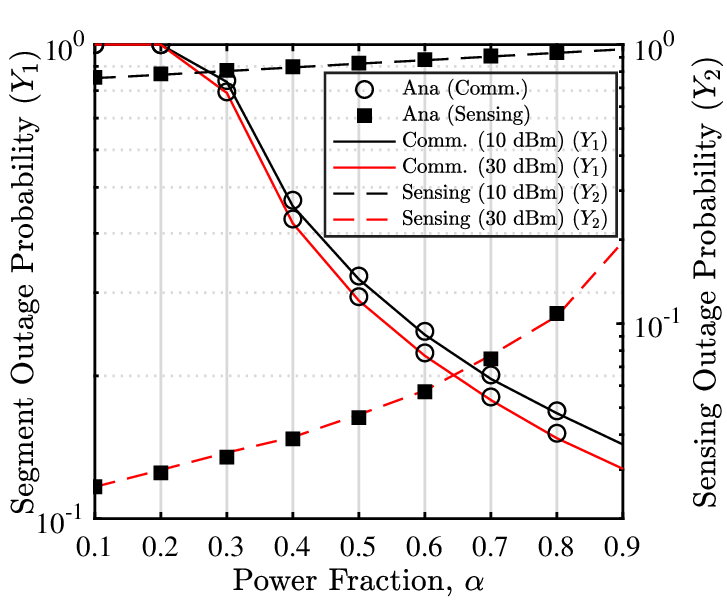}
    \caption{Segment and sensing outage analysis wrt $\alpha$. }
    \label{ISAC_final_new}
\end{figure}

In this subsection, we present the ISAC performance by jointly illustrating the segment outage probability (shown on the $Y_1$-axis) and sensing outage probability (shown on the $Y_2$-axis) with respect to the power-splitting factor $\alpha$. Here, $\alpha$ represents the fraction of the total transmit power used for communication, while the remaining $1-\alpha$ is used for sensing. For a lower value of $\alpha$, most of the available power is used for sensing, leaving limited power for communication. Also, the communication part has strong interference from adjacent segments compared to sensing, this results in a segment outage close to one at a lower value of $\alpha$. As $\alpha$ increases, the communication SINR improves, thereby reducing the segment outage. On the other hand, less power becomes available for sensing, resulting in a higher sensing outage. This opposite behaviour of the communication and sensing outages is illustrated in \figurename{~\ref{ISAC_final_new}}.      
It can also be observed that, for communication, the SINR is more interference-dependent due to the inter-segment and intra-segment interference. Therefore, increasing the transmit power has a limited effect on the communication SINR. In the sensing case, intra-segment interference and interference from the adjacent sensing signals are not present. Hence, the sensing SINR is more dependent on the transmit power than on interference. This behaviour can be clearly observed in 
\figurename{~\ref{ISAC_final_new}}, where a large gap exists between the sensing outage curves for $10$ dBm and $30$ dBm transmit power. In comparison to this, the gap between the corresponding communication segment outage curves is much smaller. This shows that the communication SINR is mainly interference-dependent, whereas the sensing SINR is more power-dependent. 
\balance
\section{Conclusion}
\label{sec:conclusion}
This paper proposed a V-SWAN-assisted downlink NOMA-ISAC  EDMA framework for communication and sensing, where multiple PAs are deployed along a vertically segmented waveguide to extend the coverage capability of PASS from a conventional 2D plane to a 3D space, while simultaneously serving multiple users and performing target sensing. The signal models are formulated along with SINR expressions. Analytical expressions are formulated for multi-user, multi-PA systems using statistical characterizations of the ground communication, aerial communication, and sensing channels. Based on these characterizations, the segment outage probability and sensing outage probability are formulated and evaluated using numerical integration methods. Monte Carlo simulations validated the accuracy of the derived analytical expressions and demonstrated the performance gain of the proposed framework over conventional state-of-the-art methods. The results further revealed that the communication and sensing trade-off is significantly influenced by the user and target distributions, power allocation, waveguide attenuation, and sensing-related parameters, as these factors affect the effective channel gain, interference level, and sensing performance.

\bibliographystyle{IEEEtran}
\bibliography{references}
\end{document}